\documentclass[11pt]{article}
\usepackage[dvips]{graphicx}
\usepackage{psfrag}
\usepackage{amssymb}
\usepackage{amsmath}
\usepackage{amscd}
\usepackage{latexsym}

\catcode`@=11
\makeatletter\renewcommand{\section}{\@startsection
{section}{1}{\z@}{-3.5ex plus -1ex minus
    -.2ex}{2.3ex plus .2ex}{\large\bf }}

\makeatletter\renewcommand{\subsection}{\@startsection{subsection}{2}{\z@}{-3.25ex
plus -1ex minus
   -.2ex}{1.5ex plus .2ex}{\bf }}

\numberwithin{equation}{section}
\newcounter{saveeqn}

\catcode`@=12

\def\a{\alpha}
\def\b{\beta}
\def\ga{\gamma}
\def\Ga{\Gamma}
\def\de{\delta}

\def\vk{\varkappa}
\def\r{\rho}
\def\s{\sigma}

\def\vp{\varphi}
\def\om{\omega}
\newcommand{\C}{\mathbb C}
\newcommand{\R}{\mathbb R}
\newcommand{\N}{\mathbb N}
\newcommand{\Hcal}{{\cal H}}
\newcommand{\Acal}{{\cal A}}
\newcommand{\Ecal}{{\cal E}}
\newcommand{\Rcal}{{\cal R}}
\newcommand{\Fcal}{{\cal F}}
\newcommand{\Dcal}{{\cal D}}
\newcommand{\gfrak}{{\mathfrak g}}
\newcommand{\mfrak}{{\mathfrak m}}
\newcommand{\hfrak}{{\mathfrak h}}
\newcommand{\ph}{{\widehat{\smash{\phi}}}}
\newcommand{\vph}{{\widehat{\smash{\varphi}}}}
\newcommand{\ch}{{\widehat{\smash{\chi}}}}

\def\im{\textrm{i}}
\def\ep{\textrm{e}}

\def\diff{\textrm{d}}
\def\tr{\textrm{tr}}
\def\sfrac#1#2{{\textstyle\frac{#1}{#2}}}
\def\+{\dagger}
\def\={\ =\ }
\def\und{\qquad\textrm{and}\qquad}
\def\and{\quad\textrm{and}\quad}

\def\am{\mathrm{am}}        
\def\sn{\mathrm{sn}}        
\def\cn{\mathrm{cn}}        
\def\dn{\mathrm{dn}}        

\begin{document}

\begin{titlepage}
\setcounter{page}{0}
\begin{flushright}
ITP--UH--10/10\\
\end{flushright}

\vskip 2.0cm

\begin{center}

{\Large\bf
Yang-Mills instantons and dyons on homogeneous $G_2$-manifolds
}

\vspace{12mm}

{\Large Irina~Bauer${}^\+$,\ Tatiana~A.~Ivanova${}^*$,\ 
Olaf~Lechtenfeld${}^{\+\times}$ \ and \ Felix~Lubbe${}^\+$
}
\\[8mm]
\noindent ${}^\dagger${\em 
Institut f\"ur Theoretische Physik,
Leibniz Universit\"at Hannover \\
Appelstra\ss{}e 2, 30167 Hannover, Germany }\\
{Emails: Irina.Bauer, Olaf.Lechtenfeld, Felix.Lubbe@itp.uni-hannover.de}
\\[8mm]
\noindent ${}^\times${\em
Centre for Quantum Engineering and Space-Time Research\\
Leibniz Universit\"at Hannover \\
Welfengarten 1, 30167 Hannover, Germany }\\
{URL: http://www.questhannover.de/}
\\[8mm]
\noindent ${}^*${\em 
Bogoliubov Laboratory of Theoretical Physics, JINR\\
141980 Dubna, Moscow Region, Russia}\\
{Email: ita@theor.jinr.ru}

\vspace{12mm}

\begin{abstract}
\noindent
We consider Lie$G$-valued Yang-Mills fields on the space $\R{\times}G/H$,
where $G/H$ is a compact nearly K\"ahler six-dimensional homogeneous space,
and the manifold $\R{\times}G/H$ carries a $G_2$-structure. After imposing
a general $G$-invariance condition, Yang-Mills theory with torsion on
$\R{\times}G/H$ is reduced to Newtonian mechanics of a particle moving in
$\R^6$, $\R^4$ or $\R^2$ under the influence of an inverted double-well-type 
potential for the cases $G/H=$ SU(3)/U(1)${\times}$U(1), 
Sp(2)/Sp(1)${\times}$U(1) or $G_2$/SU(3), respectively.
We analyze all critical points and present analytical and numerical
kink- and bounce-type solutions, which yield $G$-invariant instanton 
configurations on those cosets. Periodic solutions on $S^1{\times}G/H$ 
and dyons on $\im\R{\times}G/H$ are also given.
\end{abstract}

\end{center}
\end{titlepage}

\section{Introduction and summary}

Interest in Yang-Mills theories in dimensions greater than
four grew essentially after the discovery of superstring theory,
which contains supersymmetric Yang-Mills in the low-energy limit
in the presence of D-branes as well as in the heterotic case.
In particular, heterotic strings yield $d{=}10$ heterotic supergravity
interacting with the ${\cal N}{=}1$ supersymmetric Yang-Mills
multiplet~\cite{GSW}. Supersymmetry-preserving compactifications on
spacetimes $M_{10{-}d}\times X^d$ with further reduction to $M_{10{-}d}$
impose the first-order BPS-type gauge equations which are a generalization
of the Yang-Mills anti-self-duality equations in $d{=}4$ to higher-dimensional
manifolds with special holonomy. Such equations in $d{>}4$ dimensions
were first introduced in~\cite{Corrigan:1982th} and further considered e.g.\
in~\cite{Ward84}-\cite{Popov}. Some of their solutions were found e.g.\
in~\cite{group1}-\cite{HILP}.

Initial choices for the internal manifold $X^6$ in string theory were K\"ahler
coset spaces and Calabi-Yau manifolds, as well as manifolds with exceptional
holonomy group $G_2$ for $d{=}7$ and Spin(7) for $d{=}8$. However,
it was realized recently that the internal manifold should allow
non-trivial $p$-form fluxes whose back reaction deforms its geometry.
In particular, a three-form flux background implies a nonzero torsion
whose components are given by the structure constants of the holonomy group,
$T^a_{bc}=\vk\,f^a_{bc}$, with a real parameter~$\vk$.
String vacua with $p$-form fields along the extra dimensions (`flux
compactifications') have been intensively studied in recent years (see
e.g.~\cite{group3} for reviews and references).
Flux compactifications have been investigated primarily for type II strings
and to a lesser extent in the heterotic theories, despite their long history
\cite{hetold}. The number of torsionful geometries that can serve as a
background for heterotic string compactifications seems rather limited.
Among them there are six-dimensional nilmanifolds, solvmanifolds,
nearly K\"ahler and nearly Calabi-Yau coset spaces. The last two kinds
of manifolds carry a natural almost complex structure which is not integrable
(for their geometry see e.g.~\cite{But, group4} and references therein).

In the present paper, we solve the torsionful Yang-Mills equations on
$G_2$-manifolds of topology $\R{\times}X^6$ with nearly K\"ahler cosets $X^6$.
The allowed gauge bundle is restricted by the $G_2$-instanton 
equations~\cite{DT}. For each coset~$X^6=G/H$, we parametrize the general 
$G$-invariant connection by a set of complex scalars~$\phi_i$, which depend 
on the coordinate~$\tau$ of the $\R$~factor. The Yang-Mills equations then
descend to Newton's equations for the coordinates~$\phi_i(\tau)$ of a point
particle under the influence of an inverted double-well-type potential,
whose shape depends on~$\vk$. For this potential we derive the critical points
of zero energy, which correspond to the $\tau{\to}{\pm}\infty$ asymptotic
configurations of the finite-action Yang-Mills solutions. We then present
a variety of zero-energy solutions~$\phi_i(\tau)$, of kink and of bounce type,
analytically as well as numerically. The kinks translate to instantons
for the gauge fields.

Furthermore, by replacing the factor~$\R$ with $S^1$, we obtain
periodic solutions with a sphaleron interpretation. Finally, in the
Lorentzian case $\im\R{\times}G/H$, the double-well-type potential gets
flipped back, and there exist bounce solutions with a dyonic interpretation, 
some of which have finite action. The different
types of finite-action Yang-Mills solutions on $\R{\times}G/H$ or 
$\im\R{\times}G/H$ occur in the following ranges of the parameter~$\vk$:
\begin{center}
\begin{tabular}{|c|cccccc|}
\hline
$\vk\in$ & $(-\infty,-3)$ & $(-3,+1)$ & $(+1,+3)$ & 
$(+3,+5)$ & $(+5,+9)$ & $(+9,+\infty)$\\
\hline
Euclidean & bounces & instantons & instantons & bounces & --- & --- \\
Lorentzian & dyons & --- & --- & --- & dyons & dyons \\
\begin{minipage}{2cm} $\ \ V_{\R}(\textrm{Re}\phi)$ \\[2pt] \end{minipage} & 
\includegraphics[width=1.5cm]{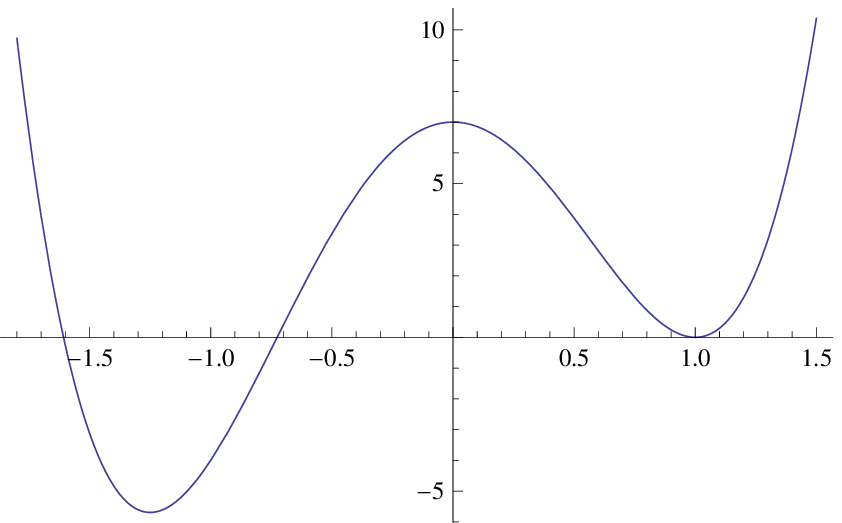} &
\includegraphics[width=1.5cm]{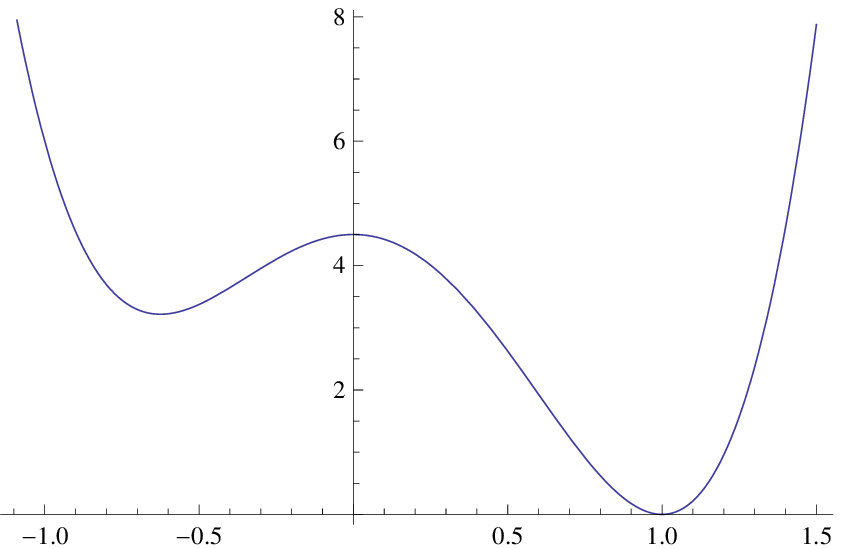} & 
\includegraphics[width=1.5cm]{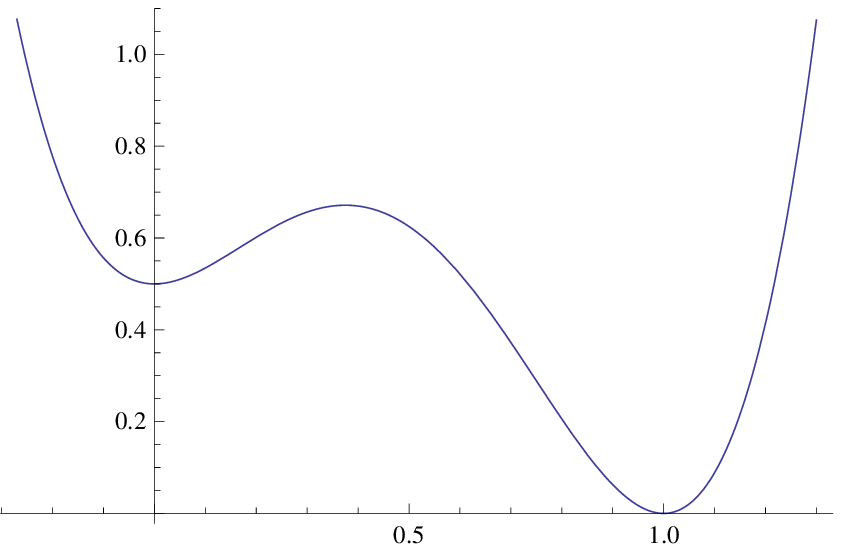} &
\includegraphics[width=1.5cm]{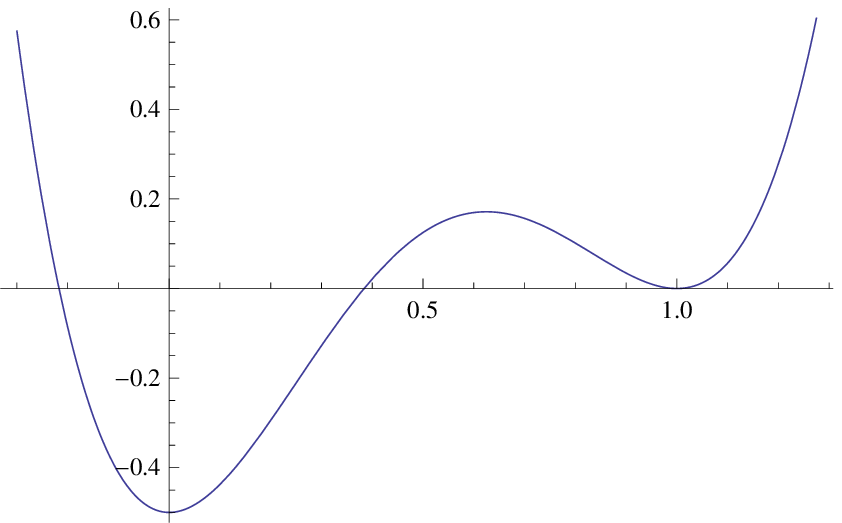} &
\includegraphics[width=1.5cm]{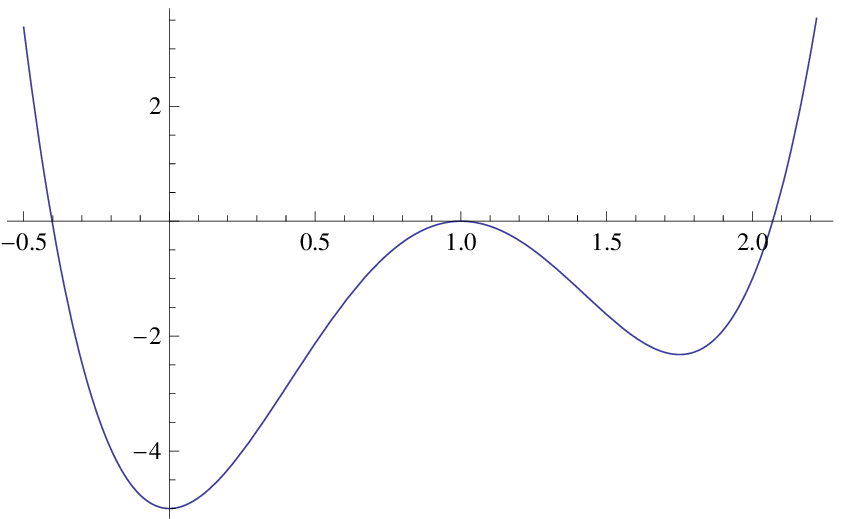} &
\includegraphics[width=1.5cm]{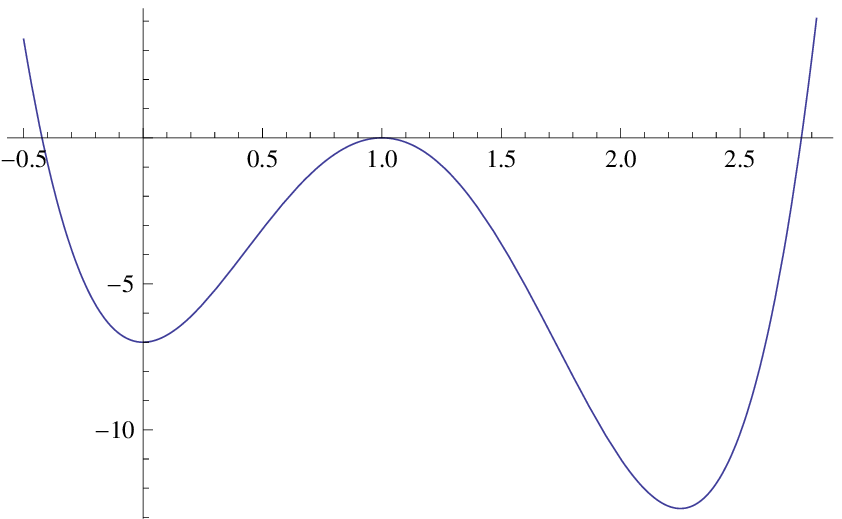} \\
\hline
\end{tabular}
\end{center}

\section{Yang-Mills fields on $\R\times G/H$}

\subsection{Yang-Mills equations with torsion}

Instantons~\cite{BPST} play an important role in  modern gauge
theories~\cite{Raj, MS}. They are nonperturbative BPS configurations
in four Euclidean dimensions solving the first-order anti-self-duality
equations and forming a subset of solutions to the full Yang-Mills
equations. In dimensions higher than four, BPS configurations can
still be found as solutions to first-order equations, known as generalized
anti-self-duality equations~\cite{Corrigan:1982th}-\cite{Bau} or
$\Sigma$-anti-self-duality~\cite{Tian, DT}. These appear in superstring
compactifications as conditions of survival of at least one
supersymmetry~\cite{GSW}. Various solutions to these first-order equations
were found e.g. in~\cite{group1}-\cite{HILP}, mostly on flat space $\R^d$
and various cosets.

The BPS-type instanton equations in $d>4$ dimensions can be introduced
as follows. Let $\Sigma$ be a $(d{-}4)$-form on a $d$-dimensional Riemannian
manifold $M$. Consider a complex vector bundle $\Ecal$ over $M$ endowed
with a connection $\Acal$. The $\Sigma$-anti-self-dual gauge equations
are defined~\cite{Tian} as the first-order equations,
\begin{equation}\label{2.1}
\ast\Fcal \= - \Sigma\wedge\Fcal\ ,
\end{equation}
on a connection $\Acal$ with the curvature 
$\Fcal =\diff\Acal +\Acal\wedge\Acal$.
Here $\ast$ is the Hodge star operator on $M$.

Differentiating (\ref{2.1}), we obtain the Yang-Mills equations with torsion,
\begin{equation}\label{2.2}
\diff\ast\Fcal + \Acal\wedge\ast\Fcal - \ast\Fcal\wedge\Acal +
\ast\Hcal\wedge\Fcal \=0\ ,
\end{equation}
where the torsion three-form $\Hcal$ is defined by the formula
\begin{equation}\label{2.3}
\ast\Hcal\ :=\ \diff\Sigma\qquad\Rightarrow\qquad 
\Hcal \= (-1)^{3(d-3)}\ast\diff\Sigma\ .
\end{equation}
The torsion term in (\ref{2.2}) naturally appears in string
theory~\cite{group3}.\footnote{For a recent discussion of heterotic string
theory with torsion see e.g.~\cite{group5}-\cite{group7} and references therein.}
If $\Sigma$ is closed, $\Hcal=0$ and (\ref{2.2}) reduce to the standard
Yang-Mills equations. The Yang-Mills equations with torsion (\ref{2.2}) are
equations of motion for the action
\begin{equation}\label{2.4}
\begin{aligned}
S&\=\int\limits_M^{}\tr\left(\Fcal\wedge\ast\Fcal\ +\
(-1)^{d-3}\Sigma\wedge\Fcal\wedge\Fcal\right)\\
&\=\int\limits_M^{}\tr\,\Bigl(\Fcal\wedge\ast\Fcal\ +\ \ast\Hcal\wedge
\bigl(\diff\Acal\wedge\Acal +\sfrac{2}{3}\Acal^3\bigr)\Bigr)\ -\
\int\limits_M^{}\diff\Bigl(\Sigma\wedge\tr\,\bigl(\Acal\wedge\diff\Acal
+\sfrac{2}{3}\Acal^3\bigr) \Bigr)\ ,
\end{aligned}
\end{equation}
where the last term is topological. In what follows we
consider the equations (\ref{2.2}) on manifolds $M=\R\times G/H$, where
$G/H$ are compact nearly K\"ahler six-dimensional homogeneous spaces.

\subsection{Coset spaces}

Consider a compact semisimple Lie group $G$ and a closed subgroup $H$ of $G$
such that $G/H$ is a reductive homogeneous space (coset space). 
Let $\{I_A\}$ with $A{=}1,\ldots,\,$dim\,$G$ be the generators of the
Lie group $G$ with structure constants $f^A_{BC}$ given by the commutation
relations
\begin{equation}\label{2.5}
[I_A, I_B]\=f^C_{AB}\, I_C \ .
\end{equation}
We normalize the generators such that the Killing-Cartan metric
on the Lie algebra $\gfrak$ of $G$ coincides with the Kronecker symbol,
\begin{equation}\label{2.6}
g_{AB}\=f^C_{AD}\,f^D_{CB}\=\de_{AB} \ .
\end{equation}
More general left-invariant metrics can be obtained by rescaling the
generators.

The Lie algebra $\gfrak$ of $G$ can be decomposed as
${\gfrak}={\hfrak}\oplus{\mfrak}$, where $\mfrak$ is the orthogonal complement
of the Lie algebra $\hfrak$ of $H$ in $\gfrak$. Then, the generators of $G$ can
be divided into two sets, $\{I_A\}=\{I_a\}\cup\{I_i\}$, where $\{I_i\}$ are
the generators of $H$ with 
$i,j,\ldots=\textrm{dim}\,G{-}\textrm{dim}\,H{+}1,\ldots,\textrm{dim}\,G$,
and $\{I_a\}$ span the subspace $\mfrak$ of $\gfrak$ with $a,b,\ldots=
1,\ldots,\,$dim\,$G{-}$dim\,$H$.
For reductive homogeneous spaces we have the following commutation relations:
\begin{equation}\label{2.7}
[I_i, I_j]\=f^k_{ij}\, I_k \ ,\qquad [I_i, I_a]\=f^b_{ia}\, I_b\und
[I_a, I_b]\=f^i_{ab}\, I_i + f^c_{ab}\, I_c\ .
\end{equation}
For the metric (\ref{2.6}) on $\gfrak$ we have
\begin{equation}\label{2.8}
g_{ab}\=2f^i_{ad}f^d_{ib} + f^c_{ad}f^d_{cb}\=\de_{ab}\ ,
\end{equation}
\begin{equation}\label{2.9}
g_{ij}\=f^k_{il}f^l_{kj} + f^b_{ia}f^a_{bj}\=\de_{ij} \und g_{ia}\=0\ .
\end{equation}

\subsection{Torsionful spin connection on $G/H$}
The metric (\ref{2.8}) on $\mathfrak{m}$ lifts to a $G$-invariant
metric on $G/H$.  A local expression for this can be obtained by introducing
an orthonormal frame as follows.  The basis elements $I_A$
of the Lie algebra $\mathfrak{g}$ can be represented by left-invariant vector
fields $\hat E_A$ on the Lie group $G$, and the dual basis $\hat{e}^A$ is a set
of left-invariant one-forms.  The space $G/H$ consists of left cosets $gH$ and
the natural projection $g\mapsto gH$ is denoted $\pi:G\rightarrow G/H$.
Over a small contractible open subset $U$ of $G/H$, one can choose a map
$L:U\rightarrow G$ such that $\pi\circ L$ is the identity, i.e. $L$ is a local
section of the principal bundle $G\rightarrow G/H$.  The pull-backs of
$\hat{e}^A$ by $L$ are denoted $e^A$.  Among these, the $e^a$ form an 
orthonormal frame for $T^* (G/H)$ over $U$, and for the remaining forms
we can write $e^i=e^i_ae^a$ with real functions $e^i_a$. 
The dual frame for $T(G/H)$ will be denoted $E_a$.  
By the group action we can transport $e^a$ and $E_a$ from inside $U$ 
to everywhere in $G/H$. The forms $e^A$ obey the Maurer-Cartan equations,
\begin{equation}\label{2.10}
\diff e^a \=
- f^a_{ib}\, e^i\wedge e^b -\sfrac{1}{2}\, f^a_{bc}\, e^b\wedge e^c
\und
\diff e^i \=
- \sfrac12\,f^i_{bc}\, e^b\wedge e^c -\sfrac12\, f^i_{jk}\, e^j\wedge e^k \ .
\end{equation}
The local expression for the $G$-invariant metric then is
\begin{equation}\label{2.11}
g_{\scriptscriptstyle G/H}^{} \= \delta_{ab}e^a e^b\ .
\end{equation}

Recall that a linear connection is a matrix of one-forms $\Ga = (\Ga^a_b) =
(\Ga^a_{cb}e^c)$. The connection is metric compatible if $g_{ac}\Ga^c_b$
is anti-symmetric, and its torsion is a vector of two-forms $T^a$ determined
by the structure equations
\begin{equation}\label{2.12}
\diff e^a + \Ga^a_{b}\wedge e^b\=T^a\=\sfrac{1}{2}\,T^a_{bc}\,e^b\wedge e^c\ .
\end{equation}
We choose the torsion tensor components on $G/H$ proportional to the
structure constants $f^a_{bc}$,
\begin{equation}\label{2.14}
T^a_{bc}\=\vk\,f^a_{bc}\ ,
\end{equation}
where $\vk$ is an arbitrary real parameter.
Then the torsionful spin connection on $G/H$ becomes
\begin{equation}\label{2.15}
\Ga^a_{b}\=f^a_{ib}e^i+\sfrac12\,(\vk{+}1)\,f^a_{cb}\,e^c\ =:\ \Ga^a_{cb}e^c\ .
\end{equation}

\subsection{Yang-Mills equations on $\R\times G/H$}

Consider the space $\R\times G/H$ with a coordinate $\tau$ on $\R$,
a one-form $e^0:=\diff\tau$ and the Euclidean metric
\begin{equation}\label{2.17}
g \= (e^0)^2\ +\ \de_{ab}\,e^a e^b\ .
\end{equation}
The torsionful spin connection $\Ga$ on $\R\times G/H$ is given 
by~(\ref{2.15}), with 
\begin{equation}\label{2.18}
\Ga^a_{cb}\=e^i_c\,f^a_{ib}+\sfrac{1}{2}\,(\vk{+}1)\,f^a_{cb} \und
\Ga^0_{0b}=\Ga^a_{0b}=\Ga^0_{cb}=0\ .
\end{equation}
For our choice of the metric, $g_{ab}=\de_{ab}$, we can pull down indices
in (\ref{2.14}) and introduce the three-form
\begin{equation}\label{2.19}
\Hcal\=\sfrac{1}{3!}\,T_{abc}\,e^a\wedge e^b\wedge e^c\=\sfrac{1}{6}\,\vk
f_{abc}\,e^a\wedge e^b\wedge e^c 
\qquad\Longrightarrow\qquad
\Hcal_{abc}=T_{abc}=\vk f_{abc}\ .
\end{equation}

Consider the trivial principal bundle $P(\R{\times}G/H, G)=
(\R{\times}G/H){\times}G$ over $\R{\times}G/H$ with the structure group $G$,
the associated trivial complex vector bundle $\Ecal$ over $\R{\times}G/H$
and a $\gfrak$-valued connection one-form $\Acal$ on $\Ecal$ with the
curvature $\Fcal =\diff\Acal +\Acal\wedge\Acal$. In the basis of one-forms
$\{e^0, e^a\}$ on $\R{\times}G/H$, we have
\begin{equation}\label{2.21}
\Acal \=\Acal_0e^0 + \Acal_a e^a \und
\Fcal \=\Fcal_{0a}\,e^0\wedge e^a + \sfrac12\, \Fcal_{ab}\,e^a \wedge e^b\ .
\end{equation}
In the following we choose a `temporal' gauge in which
$\Acal_0\equiv\Acal_\tau =0$.

The Yang-Mills equations with torsion (\ref{2.2}) on $\R{\times}G/H$ are
equivalent to
\begin{eqnarray}\label{2.22}
E_a\Fcal^{a0}+ \Ga^a_{ab}\Fcal^{b0}+[\Acal_a,\Fcal^{a0}] &=& 0\ , \\[6pt]
\label{2.23}
E_0\Fcal^{0b}+E_a\Fcal^{ab}+ \Ga^d_{da}\Fcal^{ab}+ \Ga^b_{cd}\Fcal^{cd}
+[\Acal_a,\Fcal^{ab}] &=& 0\ ,
\end{eqnarray}
where we used (\ref{2.18}) and (\ref{2.19}) 
and the gauge $\Acal_0=0$ with $E_0=\diff/\diff\tau$. 
Note that these equations also follow from the action
functional (\ref{2.4}) with $\Hcal$ given in (\ref{2.19}).

\subsection{$G$-invariant gauge fields}

Let us associate our complex vector bundle $\Ecal\to\R\times G/H$ 
with the adjoint representation adj($G$) of the structure group~$G$.
Then the generators of $G$ are realized as dim\,$G\times$dim\,$G$ matrices
\begin{equation}\label{2.24}
I_i\=\bigl(I_{iB}^A\bigr)\=\bigl(f_{iB}^A\bigr)\=
\bigl(f_{ik}^j\bigr)\oplus\bigl(f_{ib}^a\bigr) \und 
I_a\=\bigl(I_{aB}^A\bigr)\=\bigl(f_{aB}^A\bigr)\ .
\end{equation}
According to~\cite{KN} 
(see also~\cite{Kub, KZ, LPS}), $G$-invariant connections on $\Ecal$ 
are determined by linear maps $\Lambda:\mfrak\to\gfrak$ which commute 
with the adjoint action of $H$: 
\begin{equation}
\Lambda\bigl(\mbox{Ad}(h)Y\bigr)\=\mbox{Ad}(h)\,\Lambda (Y) 
\qquad\forall h\in H \and Y\in\mfrak\ . 
\end{equation}
Such a linear map is represented by a matrix $(X^B_a)$, appearing in
\begin{equation}
X_a\ :=\ \Lambda(I_a) \= X^B_a I_B \= X^i_a I_i + X^b_a I_b\ .
\end{equation}
For the cases we will consider one can always choose $X^i_a=0$. 
In local coordinates the connection is written
\begin{equation}\label{2.25}
\Acal \= e^iI_i + e^a X_a \qquad\Leftrightarrow\qquad
\Acal_a \= e^i_aI_i + X_a\ ,
\end{equation}
and its $G$-invariance imposes the condition
\begin{equation}\label{2.26}
[I_i, X_a]\=f^b_{ia}X_b \qquad\Leftrightarrow\qquad
X^b_a f^c_{bi} \= f^b_{ia} X^c_b\ .
\end{equation}

The curvature $\Fcal$ of the invariant connection (\ref{2.25}) reads
\begin{equation}\label{2.27}
\begin{array}{c}
\Fcal \=\diff\Acal +\Acal\wedge\Acal\=\dot X_ae^0\wedge e^a-
\sfrac{1}{2}\,\bigl(f_{bc}^iI_i + f_{bc}^aX_a-[X_b,X_c]\bigr)\,e^b\wedge e^c
\qquad\Leftrightarrow\\[6pt]
\Fcal_{0a}\=\dot X_a\und
\Fcal_{bc}\=-\bigl(f_{bc}^iI_i + f_{bc}^aX_a-[X_b,X_c]\bigr)\ ,
\end{array}
\end{equation}
where dots denote derivatives with respect to $\tau$. For our choice
(\ref{2.8}) and (\ref{2.9}) of the metric one can pull down all indices
in the Yang-Mills equations (\ref{2.22}) and (\ref{2.23}) as well as in
(\ref{2.18}). It is now a matter of computation to substitute (\ref{2.25})
and (\ref{2.27}) into (\ref{2.22}) and (\ref{2.23}), making use of the Jacobi
identity for the structure constants. One finds that (\ref{2.23}) is
equivalent to
\begin{equation}\label{2.28}
\ddot X_a\=\bigl(\sfrac12(\vk{+}1)f_{acd}f_{bcd}-f_{acj}f_{bcj}\bigr)X_b\ -\
\sfrac{1}{2}(\vk{+}3)f_{abc}[X_b,X_c]\ -\ \bigl[X_b, [X_b,X_a]\bigr]\ ,
\end{equation}
and (\ref{2.22}) reduces to the constraint
\begin{equation}\label{2.29}
[X_a, \dot X_a]\=0 \qquad\textrm{(sum over $a$)}
\end{equation}
on the matrices $X_a$. Note that the equations (\ref{2.28}) can also be 
obtained from the action (\ref{2.4}) reduced to a matrix-model action after
substituting (\ref{2.25}) and (\ref{2.27}) into (\ref{2.4}).
The subsidiary relation~(\ref{2.29}) is the Gau\ss{}-law constraint following
from the gauge fixing $\Acal_0=0$.

\section{Invariant gauge fields on homogeneous $G_2$-manifolds}

Here, we choose $G/H$ to be a compact six-dimensional nearly K\"ahler
coset space. Such manifolds are important examples of SU(3)-structure
manifolds used in flux compactifications of string theories (see
e.g.~\cite{group4, group7} and references therein).
Their geometry is fairly rigid and features a 3-symmetry, which
generalizes the reflection symmetry of symmetric spaces.
This allows for a very explicit description of their structure
and a complete parametrization of $G$-invariant Yang-Mills fields,
which we present in this section.

\subsection{Nearly K\"ahler six-manifolds}

An SU(3)-structure on a six-manifold is by definition a reduction of the 
structure group of the tangent bundle from SO(6) to SU(3).  
Manifolds of dimension six with SU(3)-structure admit a set of canonical 
objects, consisting of an almost complex structure $J$, 
a Riemannian metric $g$, a real two-form $\omega$ and
a complex three-form $\Omega$.  
With respect to $J$, the forms $\omega$ and $\Omega$
are of type (1,1) and (3,0), respectively, 
and there is a compatibility condition,
$g(J\cdot,\cdot)=\omega(\cdot,\cdot)$.  
With respect to the volume form $V_g$ of $g$, the forms $\omega$
and $\Omega$ are normalized so that
\begin{equation}\label{3.1}
\omega\wedge\omega\wedge\omega \= 6V_g \und
\Omega\wedge\bar{\Omega} \= -8\im V_g\ .
\end{equation}
Then, a nearly K\"ahler six-manifold is an SU(3)-structure manifold 
with the differentials
\begin{equation}\label{3.2}
\diff\omega \= 3\rho\,\mathrm{Im}\Omega \und
\diff\Omega \= 2\rho\,\omega\wedge\omega
\end{equation}
for some real non-zero constant $\rho$ 
(if $\rho$ was zero, the manifold would be Calabi-Yau).  
More generally, six-manifolds with SU(3)-structure are classified
by their intrinsic torsion~\cite{CS}, and nearly K\"ahler manifolds form one
particular intrinsic torsion class.

There are only four known examples of compact nearly K\"ahler six-manifolds,
and they are all coset spaces~\cite{But}:
\begin{equation}\label{3.3}
\mbox{SU}(3)/\mbox{U}(1){\times}\mbox{U}(1)\ ,\quad
\mbox{Sp}(2)/\mbox{Sp}(1){\times}\mbox{U}(1)\ ,\quad
G_2/\mbox{SU}(3)=S^6,\ \quad 
\mbox{SU}(2)^3/\mbox{SU}(2)=S^3\times S^3\ .
\end{equation}
Here Sp(1)${\times}$U(1) is chosen to be a non-maximal subgroup of Sp(2):
if the elements of Sp(2) are written as $2\times2$ quaternionic matrices, then
the elements of Sp(1)${\times}$U(1) have the form $\mathrm{diag}(p,q)$, with
$p\in$Sp(1) and $q\in$U(1).
Also, SU(2) is the diagonal subgroup of $\mbox{SU}(2)^3$.
These coset spaces are all 3-symmetric, because the subgroup $H$ is the
fixed point set of an automorphism $s$ of $G$ satisfying $s^3=\mathrm{Id}$
\cite{But}.

The 3-symmetry actually plays a fundamental role in defining the canonical
structures on the coset spaces.  The automorphism $s$ induces an automorphism
$S$ of the Lie algebra $\mathfrak{g}=\hfrak\oplus\mfrak$ of $G$ which acts
trivially on $\hfrak$ and non-trivially on $\mfrak$; one can define a map
\begin{equation}\label{3.4}
J:\mathfrak{m}\rightarrow\mathfrak{m} \qquad\textrm{by}\qquad
S|_\mathfrak{m} \= -\sfrac{1}{2} + \sfrac{\sqrt{3}}{2} J \=
\exp\left( \sfrac{2\pi}{3}\, J \right)\ .
\end{equation}
The map $J$ satisfies $J^2=-1$ and provides the almost complex structure 
on $G/H$.
The components $J^a_b$ of the almost complex structure $J$ are defined via
$J(I_b)=J^a_bI_a$. Local expressions for the $G$-invariant metric, almost
complex structure, and the two-form $\omega$ on a nearly K\"ahler space $G/H$
in an orthonormal frame $\{e^a\}$ are
\begin{equation}\label{3.5}
g \= \delta_{ab}e^a e^b\ , \qquad
J \= J_{a}^b e^a E_b \und
\omega \= \sfrac{1}{2}J_{ab} e^a\wedge e^b\ .
\end{equation}

One can also obtain a local expression for (3,0)-form $\Omega$ by using
(\ref{3.2}) and the Maurer-Cartan equations. From
(\ref{2.10}) one can compute $\diff\omega$ and hence $\ast\diff\omega$:
\begin{equation}\label{3.6}
\diff\omega \= -\sfrac{1}{2}\,\tilde{f}_{abc}\, e^a\wedge e^b\wedge e^c \und
\ast \diff\omega \= \sfrac{1}{2}\,f_{abc}\, e^a\wedge e^b\wedge e^c\ ,
\end{equation}
where
\begin{equation}\label{3.7}
\tilde{f}_{abc}\ :=\ f_{abd}J_{dc}
\end{equation}
are the components of a totally antisymmetric tensor on a nearly K\"ahler 
six-manifold in the list (\ref{3.3}). 
The structure constants on nearly K\"ahler cosets obey the identities
\begin{equation}\label{3.8}
f_{aci}f_{bci} \= f_{acd}f_{bcd} \= \sfrac{1}{3}\,\delta_{ab}\ ,
\end{equation}
\begin{equation}\label{3.9}
J_{cd}f_{adi} \= J_{ad}f_{cdi} \und
J_{ab} f_{abi} \= 0\ .
\end{equation}
{}From the normalization (\ref{3.1}) and (\ref{3.8}) we compute that
\begin{equation}
||\omega||^2\ :=\ \omega_{ab}\omega_{ab}\=3 \und 
||\mbox{Im}\,\Omega||^2\ :=\
(\mbox{Im}\,\Omega)_{abc}(\mbox{Im}\,\Omega)_{abc}\=4\ .
\end{equation}
So it must be that
\begin{equation}\label{3.10}
\mathrm{Im}\Omega \=
-\sfrac{1}{\sqrt{3}}\,\tilde{f}_{abc}\,e^a\wedge e^b\wedge e^c\ , \qquad
\mathrm{Re}\Omega \= - \sfrac{1}{\sqrt{3}}\, f_{abc}\,
e^a\wedge e^b\wedge e^c \und \r\=\sfrac{1}{2\sqrt{3}}\ .
\end{equation}

Note that on all four nearly K\"ahler coset spaces (\ref{3.3}) one can choose
the non-vanishing structure constants such that
\begin{equation}\label{3.11}
\{f_{abc}\}:\quad f_{135}=f_{425}=f_{416}=f_{326}=-\sfrac{1}{2\sqrt{3}}
\end{equation}
and therefore
\begin{equation}\label{3.12}
\{\tilde f_{abc}\}:\quad \tilde f_{136}=\tilde f_{426}=\tilde f_{145}=
\tilde f_{235}=-\sfrac{1}{2\sqrt{3}}
\end{equation}
for $J$ such that
\begin{equation}\label{3.13}
\om\=\sfrac12J_{ab}\,e^a\wedge e^b\=
e^1\wedge e^2+e^3\wedge e^4+e^5\wedge e^6\ .
\end{equation}
Then we have
\begin{equation}\label{3.14}
\Omega\={\rm Re}\,\Omega + \im\, {\rm Im}\,\Omega
\= e^{135}{+}e^{425}{+}e^{416}{+}e^{326}+
\im (e^{136}{+}e^{426}{+}e^{145}{+}e^{235})
\ =:\ \Theta^1\wedge\Theta^2\wedge\Theta^3\ ,
\end{equation}
where $e^{abc}\equiv e^a\wedge e^b\wedge e^c$ and
\begin{equation}\label{3.15}
\Theta^1 := e^1 +\im e^2\ ,\qquad 
\Theta^2 := e^3 +\im e^4 \und
\Theta^3 := e^5 +\im e^6
\end{equation}
are forms of type (1,0) with respect to $J$.

\subsection{Yang-Mills equations and action functional}

In the previous subsection we described the geometry of nearly K\"ahler
six-manifolds. Now we would like to consider the Yang-Mills theory on
seven-manifolds $\R{\times}G/H$, where $G/H$ is a nearly K\"ahler coset space.
Note that on such manifolds
\begin{equation}\label{3.16}
M\=\R\times G/H
\end{equation}
one can introduce three-forms
\begin{equation}\label{3.17}
\Sigma\=e^0\wedge \om\ +\ {\rm Im}\,\Omega\ ,
\end{equation}
and
\begin{equation}\label{3.18}
\Sigma'\=e^0\wedge \om\ +\ {\rm Re}\,\Omega\ .
\end{equation}
Each of the two, $\Sigma$ as well as $\Sigma'$, defines a $G_2$-structure on
$\R{\times}G/H$, i.e.\ a reduction of the holonomy group SO(7) to a
subgroup $G_2\!\subset\,$SO(7). From (\ref{3.17}) and (\ref{3.18}) one sees
that both $G_2$-structures are induced from the SU(3)-structure on $G/H$.

On the seven-manifold (\ref{3.16}), the matrix equations (\ref{2.28}) and
(\ref{2.29}) simplify to
\begin{equation}\label{3.19}
\ddot X_a\=\sfrac16(\vk{-}1)X_a\ -\ \sfrac12(\vk{+}3)f_{abc}[X_b,X_c]\ -\
\bigl[X_b,[X_b, X_a]\bigr]\ ,
\end{equation}
\begin{equation}\label{3.20}
[X_a,\dot X_a]\=0 \qquad\textrm{(sum over $a$)}
\end{equation}
after using the identities (\ref{3.8}). We notice that the equations
(\ref{3.19}) and (\ref{3.20}) are the equation of motion and the Gau\ss{} 
constraint for the action
\begin{equation}\label{3.21}
S\=-\sfrac{1}{4}\,\int_{\R\times G/H}\tr\left(\Fcal\wedge\ast\Fcal\ +\
\frac{\vk}{3}\,e^0\wedge\om\wedge\Fcal\wedge\Fcal\right)\ .
\end{equation}
Substituting (\ref{2.25}) and (\ref{2.27}) into (\ref{3.21}) and imposing
the gauge $\Acal_0=0$, we obtain
\begin{equation}\label{3.22}
\begin{array}{r}
S\=-\sfrac{1}{4}\,\mbox{Vol}(G{/}H)\int\!\!\diff\tau\;\tr\Bigl(
\dot X_a\dot X_a\ -\ \sfrac{1}{6}(\vk{-}3)f_{iab}f_{jab}I_iI_j\ +\
\sfrac{1}{6}(\vk{-}1)X_aX_a \quad \\[6pt]
-\ \sfrac{1}{3}(\vk{+}3)f_{abc}X_a[X_b,X_c]\ +\
\sfrac{1}{2}[X_b,X_c][X_b,X_c]\Bigr)\ .
\end{array}
\end{equation}
The Euler-Lagrange equations for this matrix-model action are (\ref{3.19}).

\subsection{Solution of the $G$-invariance condition}

The $G$-invariance condition (\ref{2.26}),
\begin{equation} \label{Ginv}
[I_i, X_a]\=f^b_{ia}X_b \qquad\textrm{for}\qquad
X_a \= X_a^b I_b \ \in \textrm{Lie}(G){-}\textrm{Lie}(H)\ ,
\end{equation}
says that the $X_a$ must transform in the six-dimensional 
representation~$\Rcal$ of $H$ which arises in the decomposition~(\ref{2.24}),
\begin{equation} \label{adjdecomp}
\textrm{adj}(G)\big|_H\=\textrm{adj}(H)\oplus\Rcal\ ,
\end{equation}
of the adjoint of~$G$ restricted to~$H$, 
i.e.\ $(\Rcal(I_i))_a^b = f_{ia}^b$.
It is real but reducible and decomposes into complex irreducible parts as
\begin{equation} \label{Rdecomp}
\Rcal \= \sum_{p=1}^q\Rcal_p\ \oplus\ \sum_{p=1}^q\overline{\Rcal}_p\ ,
\end{equation}
with $\sum_{p=1}^q\textrm{dim}\,\Rcal_p=3$.
This is the same $H$-representation as furnished by the~$I_a$.
Hence, for each irrep~$\Rcal_p$ one can find complex linear combinations 
$I_{\a_p}^{(p)}$ of the~$I_a$, with $\a_p=1,\ldots,\textrm{dim}\,\Rcal_p$, 
such that
\begin{equation}
[I_i\,,\,I_{\a_p}^{(p)}] \= f_{i\,\a_p}^{\b_p}\,I_{\b_p}^{(p)}
\end{equation}
close among themselves for each~$p$.
In the absence of a condition on $[X_a,X_b]$, the $X_a$ appear linearly and 
thus may always be multiplied by a common factor~$\phi_p$ inside each 
irrep~$\Rcal_p$. 
By Schur's lemma this is in fact the only freedom, i.e.
\begin{equation}
X_{\a_p}^{(p)}  \= \phi_p\,I_{\a_p}^{(p)} \qquad \textrm{with}\quad
\phi_p\in\C \and \a_p=1,\ldots,\textrm{dim}\,\Rcal_p 
\end{equation}
is the unique solution to the $G$-invariance condition inside $\Rcal_p$.
The six antihermitian matrices $X_a$ are then easily reconstructed via
\begin{equation}
\bigl\{X_a\bigr\} \= 
\Bigl\{ \sfrac12 \bigl( X_{\a_p}^{(p)}-\overline{X}_{\a_p}^{(p)} \bigr)\,,\,
\sfrac1{2\im} \bigl( X_{\a_p}^{(p)}+\overline{X}_{\a_p}^{(p)} \bigr) \Bigr\}
\end{equation}
and will depend on $q$ complex functions~$\phi_p(\tau)$.
The same holds for any smaller $G$-representation~$\Dcal$ instead of~adj($G$).

For computations, we choose a basis in $\gfrak$ such that
the first dim($\Rcal_1)$ generators $I_{\a_1}$ span $\Rcal_1$, 
the next dim($\Rcal_2$) generators $I_{\a_2}$ span $\Rcal_2$ etc., 
and the last dim($H$) generators span $\hfrak$. 
Such a basis decomposes $\Rcal$ into the said blocks.
Fusing all irreducible blocks and adj($H$) together again, 
we obtain a realization of $I_i$, $I_a$ and $X_a$ as matrices in adj($G$).
Since $G$ is the gauge group, these matrices enter in the action~(\ref{3.22}).
However, for calculations it is more convenient to take a smaller
$G$-representation~$\Dcal$. This affects only the normalization of the trace, 
\begin{equation}
\tr_{\Dcal}(I_A I_B)\=-\chi_{\Dcal}\,\delta_{AB}\ ,
\end{equation}
where the (2nd-order) Dynkin index $\chi_{\Dcal}$ depends on the representation
used. We normalize our generators such that $\chi_{\textrm{adj}(G)}=1$,
and choose $\Dcal$ in all cases (see below) such that $\chi_{\Dcal}=\sfrac16$.
With this, the constant term in the action~(\ref{3.22}) computes to
\begin{equation}
-\sfrac{1}{6}(\vk{-}3)f_{iab}f_{jab}\,\tr_{\Dcal}(I_iI_j) \= 
\sfrac{1}{36}(\vk{-}3)f_{iab}f_{iab} \= \sfrac1{18}(\vk{-}3)\ .
\end{equation}

\section{Yang-Mills fields on $\R\times\,$SU(3)/U(1)${\times}$U(1)}

\subsection{Explicit form of $X_a$ matrices}

The structure constants for SU(3) which conform with the nearly K\"ahler
structure (\ref{3.11})-(\ref{3.15}) are
\begin{equation}\label{4.1}
\begin{array}{c}
f_{135}=f_{425}=f_{416}=f_{326}=-\frac{1}{2\sqrt{3}}\ ,\\[6pt]
f_{127}=f_{347}=\frac{1}{2\sqrt{3}}\ ,\quad
f_{128}=-f_{348}=-\frac{1}{2}\and f_{567}=-\frac{1}{\sqrt{3}}\ .
\end{array}
\end{equation}
The adjoint of SU(3), restricted to U(1)${\times}$U(1), decomposes as
\begin{equation}
{\bf 8}\ (\textrm{of}\ \textrm{SU}(3)) \= 
((0,0)+(0,0))_{\textrm{adj}} +(3,1) +(-3,-1) +(3,-1) +(-3,1) +(0,2) +(0,-2)\ ,
\end{equation}
where the $\Rcal_p$ are labelled by the charges $(r,s)$ under U(1)$\times$U(1).
Obviously, we have $q{=}3$ complex parameters.
We employ the fundamental representation $\Dcal={\bf 3}$ of SU(3).
It is easy to check that indeed $\chi_{\bf 3}/\chi_{\bf 8}=1/6$.

For the generators $I_{7,8}$ of the subgroup U(1)${\times}$U(1) of
SU(3) chosen in the form
\begin{equation}\label{4.2}
I_7\=-\frac{\im}{2\sqrt{3}}\,
{\small\begin{pmatrix}0&\ 0&0\\0&\ 1&0\\0&\ 0&\!\!-1\end{pmatrix}}
\und
I_8\=\frac{\im}{6}\,
{\small\begin{pmatrix}2&0&0\\0&\!\!-1&0\\0&0&\!\!-1\end{pmatrix}}\ ,
\end{equation}
the solution to the SU(3)-invariance equation (\ref{Ginv}) then reads
\begin{equation}\label{4.3}
\begin{aligned}
X_1\=\frac{1}{2\sqrt{3}} {\small\begin{pmatrix}
0&\ 0&\!\!-\phi_1\\0&\ 0&0\\\bar\phi_1&\ 0&0\end{pmatrix}} \ ,\quad
X_3\=\frac{1}{2\sqrt{3}} {\small\begin{pmatrix}
0&\!\!-\bar\phi_2&\ 0\\\phi_2&0&\ 0\\0&0&\ 0\end{pmatrix}} \ ,\quad
X_5\=\frac{1}{2\sqrt{3}} {\small\begin{pmatrix}
0&\ 0&0\\0&\ 0&\!\!-\bar\phi_3\\0&\ \phi_3&0\end{pmatrix}} \ , \\[8pt]
X_2\=\frac{1}{2\sqrt{3}} {\small\begin{pmatrix}
0&0&\im\phi_1\\0&0&0\\\im\bar\phi_1&0&0\end{pmatrix}} \ ,\quad
X_4\=\frac{-1}{2\sqrt{3}} {\small\begin{pmatrix}
0&\im\bar\phi_2&\ 0\\\im\phi_2&0&\ 0\\0&0&\ 0\end{pmatrix}} \ ,\quad
X_6\=\frac{-1}{2\sqrt{3}} {\small\begin{pmatrix}
0\ &0&0\\0\ &0&\im\bar\phi_3\\0\ &\im\phi_3&0\end{pmatrix}} \ ,
\end{aligned}
\end{equation}
where $\phi_1, \phi_2, \phi_3$ are complex-valued functions of $\tau$.
Note that for $\phi_1=\phi_2=\phi_3=1$ from (\ref{4.3}) one obtains the
normalized basis for $\mfrak$ which yields the nearly K\"ahler structure on
SU(3)/U(1)${\times}$U(1) in the standard form (\ref{3.2}), (\ref{3.5}) and
(\ref{3.11})-(\ref{3.15}).

\subsection{Equations of motion}

Substituting (\ref{4.3}) into the action (\ref{3.22}), we obtain the Lagrangian
\begin{equation}\label{4.4}
\begin{aligned}
18\,{\cal L}&\=
6\,\bigl(|\dot\phi_1|^2{+}|\dot\phi_2|^2{+}|\dot\phi_3|^2\bigr)
\ -\ (\vk{-}3)
\ +\ (\vk{-}1)\bigl(|\phi_1|^2{+}|\phi_2|^2{+}|\phi_3|^2\bigr)
\\[4mm] &\quad
-\ (\vk{+}3)\bigl(\phi_1\phi_2\phi_3{+}\bar\phi_1\bar\phi_2\bar\phi_3\bigr)
\ +\ |\phi_1\phi_2|^2+|\phi_2\phi_3|^2+|\phi_3\phi_1|^2
+|\phi_1|^4+|\phi_2|^4+|\phi_3|^4\ ,
\end{aligned}
\end{equation}
whose quartic terms may be rewritten as
\begin{equation}
\sfrac12\bigl( |\phi_1|^4+|\phi_2|^4+|\phi_3|^4 \bigr)\ +\
\sfrac12\bigl( |\phi_1|^2+|\phi_2|^2+|\phi_3|^2 \bigr)^2\ .
\end{equation}
The equations of motion for the gauge fields on 
$\R\times\,$SU(3)/U(1)${\times}$U(1)
can be obtained by plugging (\ref{4.3}) in (\ref{3.19}) and (\ref{3.20}). 
We get
\begin{equation}\label{4.6}
\begin{array}{l}
6\,\ddot{\phi}_{1} \= 
(\vk{-}1)\,\phi_{1}\ -\ (\vk{+}3)\,\bar{\phi}_{2}\bar{\phi}_{3}\ +\
\bigl(2|\phi_{1}|^{2}+|\phi_{2}|^{2}+|\phi_{3}|^{2}\bigr)\,\phi_{1}\ ,\\[4pt]
6\,\ddot{\phi}_{2} \= 
(\vk{-}1)\,\phi_{2}\ -\ (\vk{+}3)\,\bar{\phi}_{1}\bar{\phi}_{3}\ +\
\bigl(|\phi_{1}|^{2}+2|\phi_{2}|^{2}+|\phi_{3}|^{2}\bigr)\,\phi_{2}\ ,\\[4pt]
6\,\ddot{\phi}_{3} \= 
(\vk{-}1)\,\phi_{3}\ -\ (\vk{+}3)\,\bar{\phi}_{1}\bar{\phi}_{2}\ +\
\bigl(|\phi_{1}|^{2}+|\phi_{2}|^{2}+2|\phi_{3}|^{2}\bigr)\,\phi_{3}\ ,
\end{array}
\end{equation}
as well as
\begin{equation}\label{4.7}
\phi_{1}\dot{\bar{\phi}}_{1} - \dot\phi_{1}\bar{\phi}_{1}\=
\phi_{2}\dot{\bar{\phi}}_{2} - \dot\phi_{2}\bar{\phi}_{2}\=
\phi_{3}\dot{\bar{\phi}}_{3} - \dot\phi_{3}\bar{\phi}_{3}\ .
\end{equation}
The equations (\ref{4.6}) are the Euler-Lagrange equations for the
Lagrangian (\ref{4.4}) obtained from (\ref{3.21}) after fixing the gauge
$\Acal_0=0$.

\subsection{Zero-energy critical points}

Writing the equations of motion~(\ref{4.6}) as
\begin{equation} \label{eom}
6\,\ddot\phi_i \= \frac{\partial V}{\partial\bar\phi_i}\ ,
\end{equation}
we see that they describe the motion of a particle on $\C^3$ under the
influence of the inverted quartic potential~$-V$, where
\begin{equation} \label{V3}
\begin{aligned}
V&\= -(\vk{-}3)
\ +\ (\vk{-}1)\bigl(|\phi_1|^2{+}\,|\phi_2|^2{+}\,|\phi_3|^2\bigr)
\ +\ \bigl(|\phi_1|^4{+}\,|\phi_2|^4{+}\,|\phi_3|^4\bigr)
\\[4pt] &\qquad\!
- (\vk{+}3)\bigl(\phi_1\phi_2\phi_3 + \bar\phi_1\bar\phi_2\bar\phi_3\bigr)
\ +\ |\phi_1\phi_2|^2+|\phi_2\phi_3|^2+|\phi_3\phi_1|^2\ ,
\end{aligned}
\end{equation}
or, alternatively, the dynamics of three identical particles on the
complex plane, with an external potential given by the (negative of)
the first line in~(\ref{V3}) and two- and three-body interactions in
the second line.

The potential~(\ref{V3}) is invariant under permutations of the $\phi_i$
as well as under the U(1)$\times$U(1) transformations
\begin{equation} \label{u1u1}
\bigl(\,\phi_1\,,\,\phi_2\,,\,\phi_3\bigr)\ \mapsto\ \bigl(\,
\ep^{\im\de_1}\phi_1\,,\,\ep^{\im\de_2}\phi_2\,,\,\ep^{\im\de_3}\phi_3\bigr)
\qquad\textrm{with}\quad \de_1+\de_2+\de_3 = 0 \quad\textrm{mod}\ 2\pi\ ,
\end{equation}
which include the 3-symmetry, $\phi_i\mapsto\ep^{2\pi\im/3}\phi_i$.
Such a transformation may be used to align the phases of the~$\phi_i$,
i.e.~$\arg(\phi_1)=\arg(\phi_2)=\arg(\phi_3)$.
These phases only enter in the cubic term of the potential, 
which is proportional to $\cos(\sum_i\arg\phi_i)$.
Therefore, the extrema of $V$ are attained at $\sum_i\arg\phi_i=0$ or~$\pi$,
and so, employing~(\ref{u1u1}), we may take $\phi_i\in\R$ in our search 
for them.\footnote{We thank N.~Dragon for this remark.}
Furthermore, the Noether charges of the U(1)$\times$U(1) symmetry~(\ref{u1u1})
are just the differences $\ell_i-\ell_j$ of the `angular momenta'
\begin{equation}
\ell_i\ :=\ \phi_i\dot{\bar{\phi}}_i - \dot\phi_i\bar{\phi}_i\ .
\end{equation}
Hence, the constraints~(\ref{4.7}) may be interpreted as putting these
charges to zero. Note, however, that the individual angular momenta are
not conserved, since
\begin{equation}
\dot\ell_i \= \sfrac12(\vk{+}3)\,
\bigl(\phi_1\phi_2\phi_3-\bar\phi_1\bar\phi_2\bar\phi_3\bigr)\ .
\end{equation}

Finite-action solutions~$\phi_i(\tau)$ must interpolate between 
critical points with zero potential, 
\begin{equation}
\lim_{\tau\to\pm\infty} \phi_i(\tau)\ =:\ \phi_i^\pm \und
(\phi_1^\pm,\phi_2^\pm,\phi_3^\pm)\in\bigl\{\ph\bigr\}\qquad\textrm{with}\qquad
V(\ph) \= 0 \= \diff V(\ph)\ .
\end{equation}
Modulo the symmetry~(\ref{u1u1}) and permutations,
the complete list of such critical points reads: 
\begin{center}
\begin{tabular}{|c|ccc|c|cccccc|}
\hline
type & $\ph_1$ & $\ph_2$ & $\ph_3$ & $\vk$ & 
& \multicolumn{4}{c}{eigenvalues of $V''$} & \\
\hline
A & $1$ & $1$ & $1$ & any & 
$\quad0\quad$ & $\quad0\quad$ & 
$3(\vk{+}3)$ & $2(\vk{+}4)$ & $2(\vk{+}4)$ & $5{-}\vk$ \\
\,A'& $\ep^{\im\a}$ & $\ep^{\im\a}$ & $\ep^{\im\a}$ & $-3$ &
$0$ & $0$ & $0$ & $2$ & $2$ & $8$ \\[4pt]
B & $0$ & $0$ & $0$ & $+3$ & 
$2$ & $2$ & $2$ & $2$ & $2$ & $2$ \\[3pt]
C & $0$ & $0$ & $\smash{\sqrt{1{+}\sqrt{3}}}$ & $-1{-}2\sqrt{3}$ &
$0$ & $\ga_-$ & $\ga_-$ & $\ga_+$ & $\ga_+$ & $4(1{+}\sqrt{3})$ \\
\hline
\end{tabular}
\end{center}
where $\ga_\pm=-(1{+}\sqrt{3})\pm2\sqrt{2(\smash{\sqrt{3}}{-}1)}$ 
takes the numerical values of $-0.31$ and $-5.15$.
The zero modes of~$V''$ are enforced by the symmetries; their number
indicates the dimension of the critical manifold in~$\C^3$.
A critical point is marginally stable only when $V''$ has no positive
eigenvalues. At the critical points $\dot\ell_i=0$ is guaranteed, hence
the product $\ph_1\ph_2\ph_3$ has to be real unless $\vk=-3$.
The latter value is special because all phase dependence disappears,
and the symmetry~(\ref{u1u1}) is enhanced to U(1)${}^3$. 
We will not consider this special situation (type A') further.
Appendix~A proves that the above table is complete.

\subsection{Some solutions}

Finite-action trajectories $\phi_i(\tau)$ require the conserved 
Newtonian energy to vanish,
\begin{equation}
E\ :=\ 6\,\bigl(|\dot\phi_1|^2{+}|\dot\phi_2|^2{+}|\dot\phi_3|^2\bigr)
\ -\ V(\phi_1,\phi_2,\phi_3)\ \buildrel!\over=\ 0.
\end{equation}
They can be of two types:
Either $\phi_i^+\neq\phi_i^-$ (kink), or $\phi_i^+=\phi_i^-$ (bounce).
Since this choice occurs for each value of $i=1,2,3$, mixed solutions
are possible. We now present some special cases.

\noindent
{\bf Transverse kinks at ${-}3{<}\vk{<}{+}3$. \ }
The two-dimensional type A critical manifold exists for any value of~$\vk$,
so one may try to find trajectories connecting two critical points of type~A.
As a particularly symmetric choice we wish to interpolate
\begin{equation}
(\phi_i^-)=(1\,,\ep^{2\pi\im/3},\ep^{-2\pi\im/3}) 
\qquad\longrightarrow\qquad
(\phi_i^+)=(\ep^{2\pi\im/3},\ep^{-2\pi\im/3},1)\ .
\end{equation}
The three independent conserved quantities $(E,\ell_i{-}\ell_j)$ do not
suffice to integrate the equations of motion~(\ref{4.6}), so generically
one has to resort to numerical methods. With a little effort, zero-energy
`transverse' kinks can be found in the range $\vk\in(-3,+3)$. 
We display the trajectory $(\phi_i(\tau))\in\C^3$ as three curves 
$\phi_i(\tau)\in\C$ in Fig.~1 for $\vk=-2,-1,0,+1,+2$. 
Apparently, the 3-symmetry effects a permutation since
$\phi_2(\tau)=\ep^{2\pi\im/3}\phi_1(\tau)=\ep^{-2\pi\im/3}\phi_3(\tau)$.
This relation takes care of the constraint~(\ref{4.7}).
Of course, acting with the transformations~(\ref{u1u1}) generates a
two-parameter family of such `transverse' kinks.
\begin{figure}[ht]
\centerline{
\includegraphics[width=7cm]{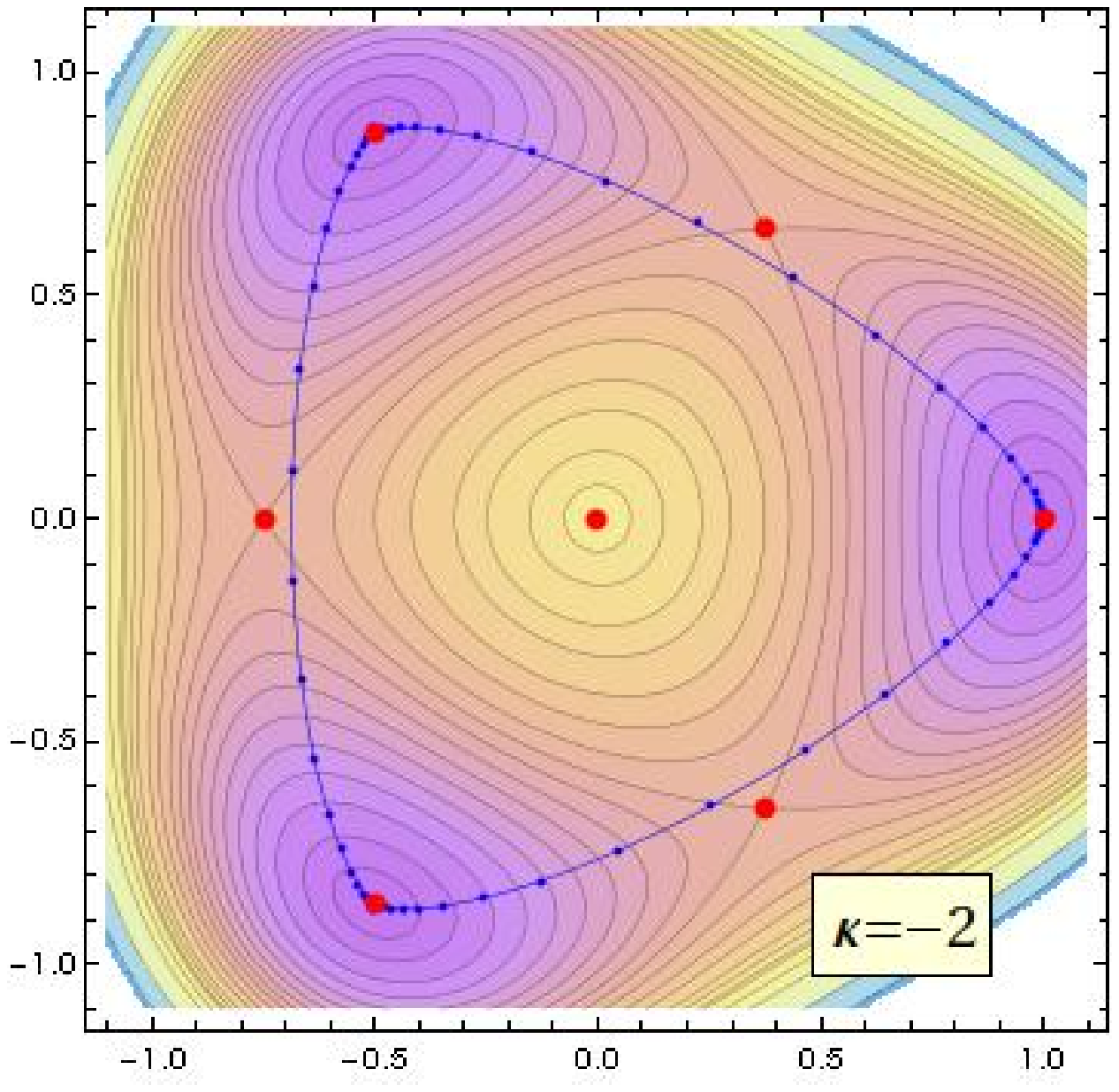}
\hfill
\includegraphics[width=7cm]{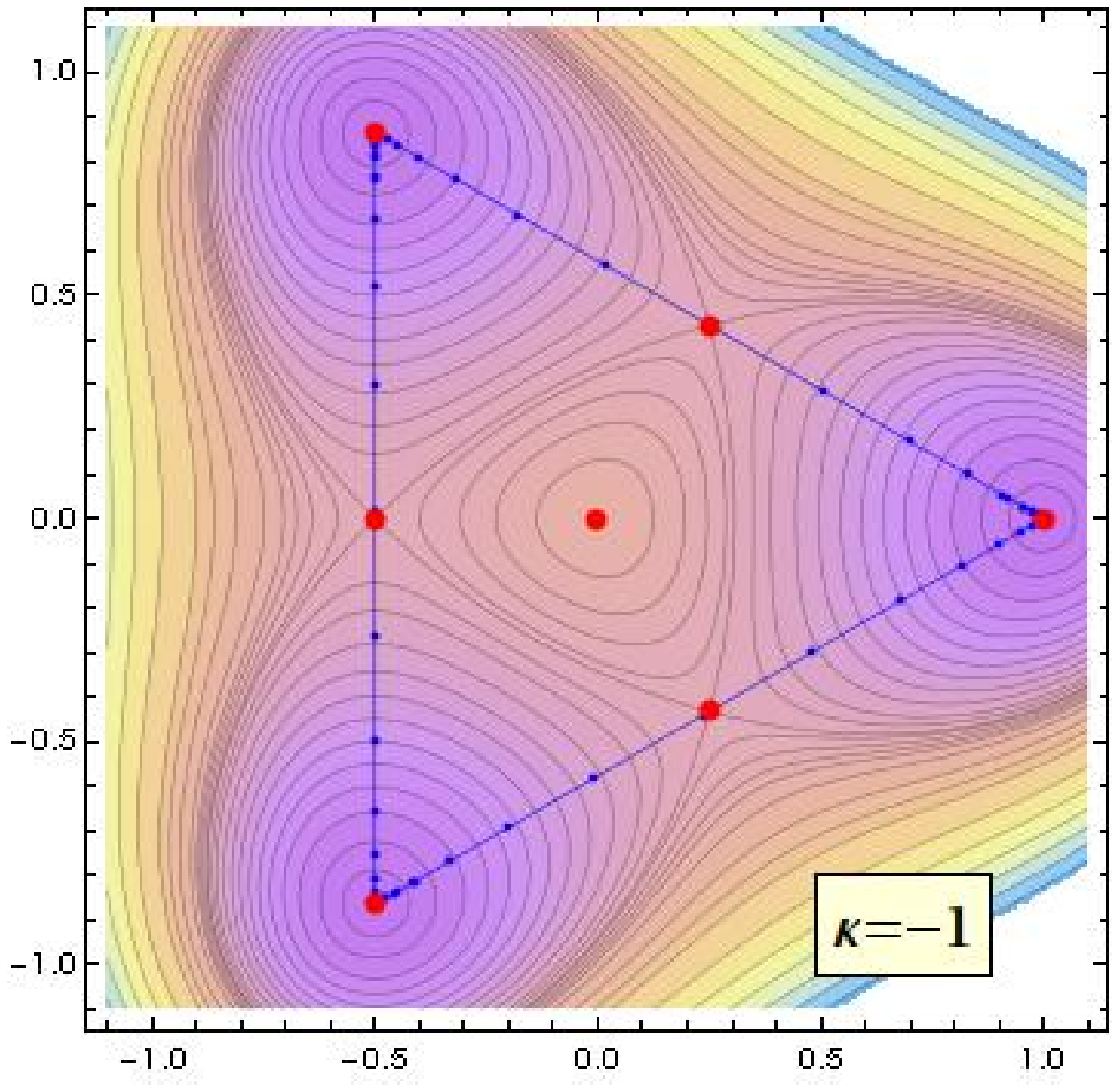}
}
\bigskip
\centerline{
\includegraphics[width=7cm]{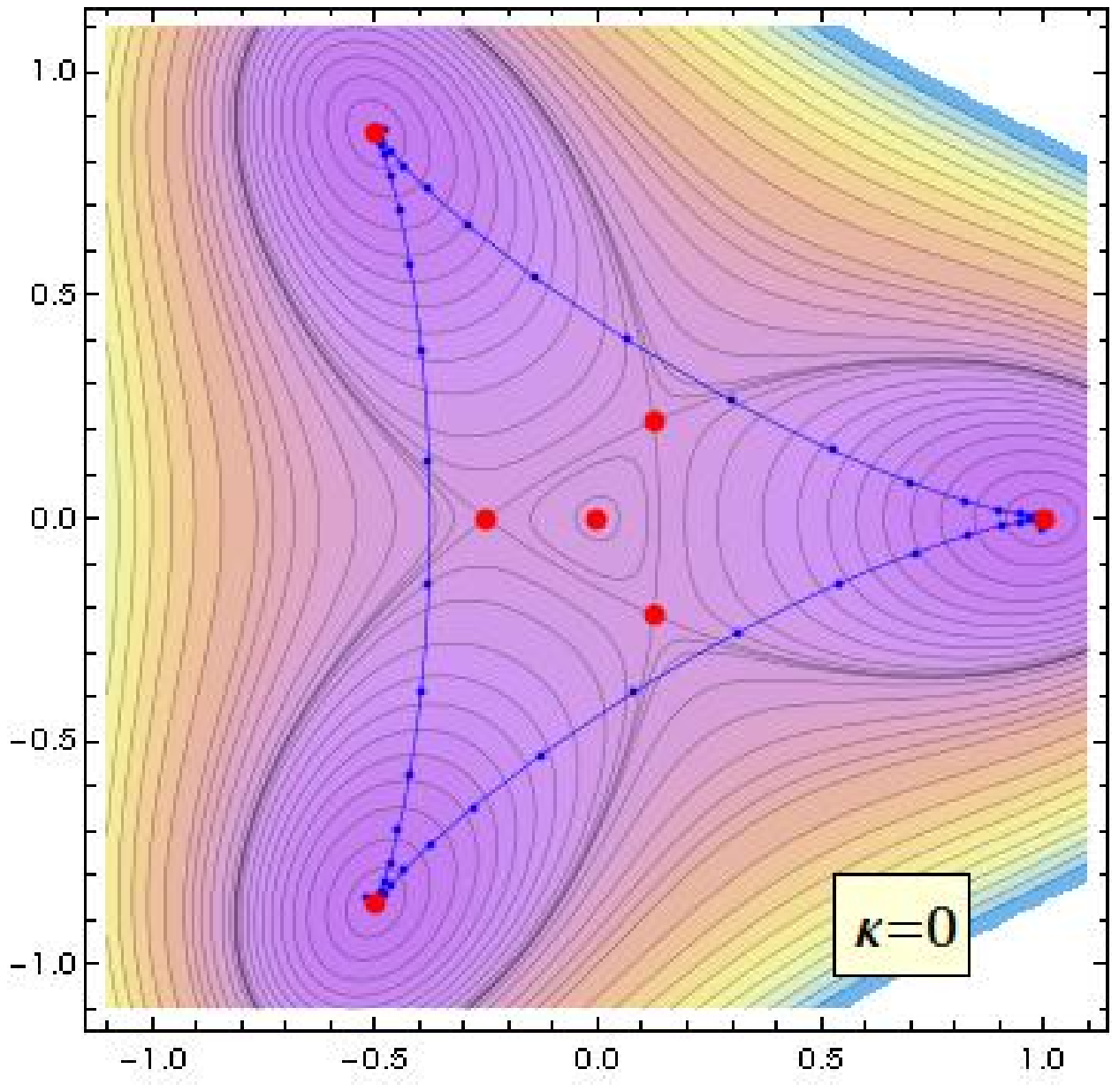}
\hfill
\includegraphics[width=7cm]{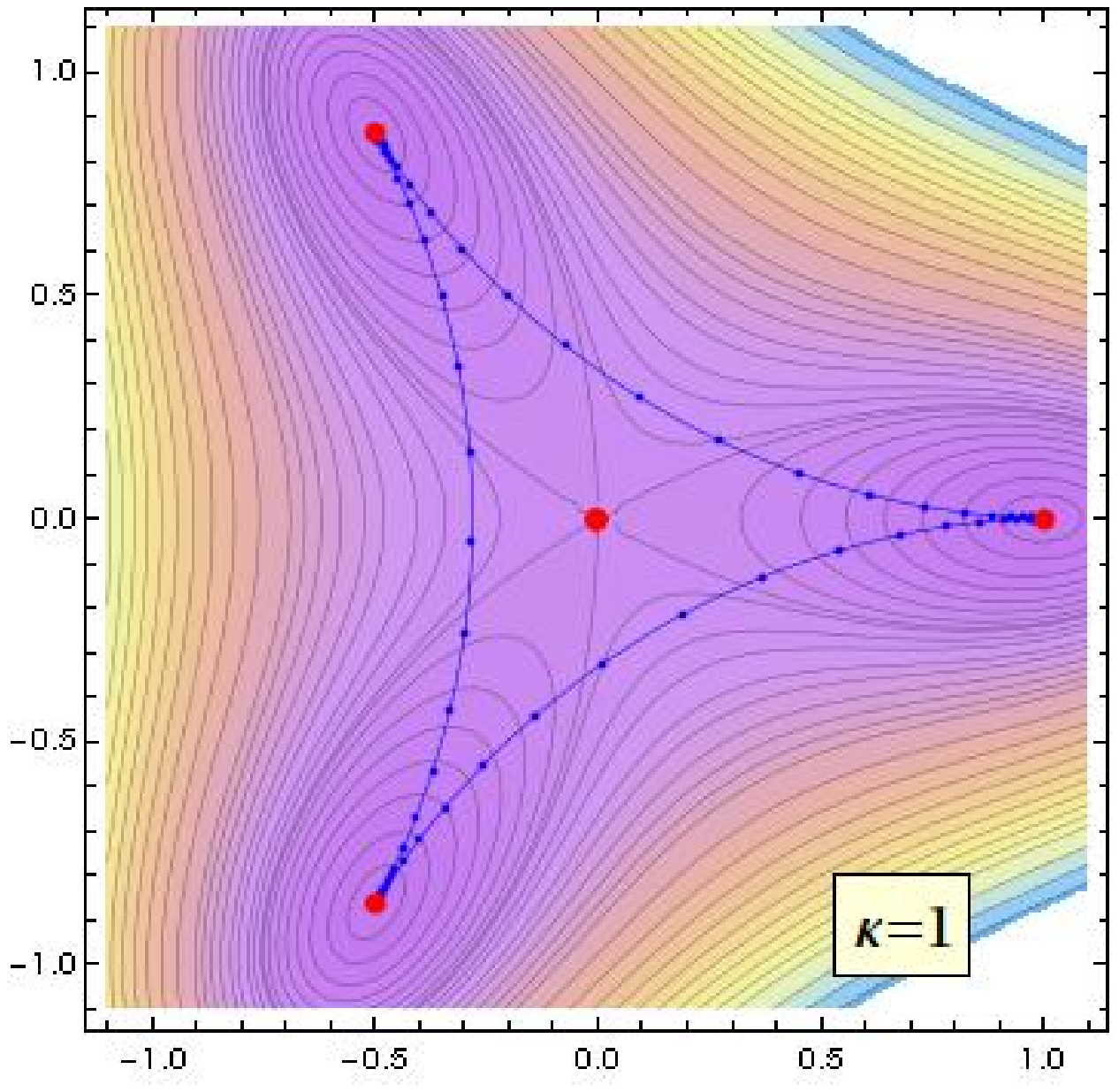}
}
\bigskip
\centerline{
\includegraphics[width=7cm]{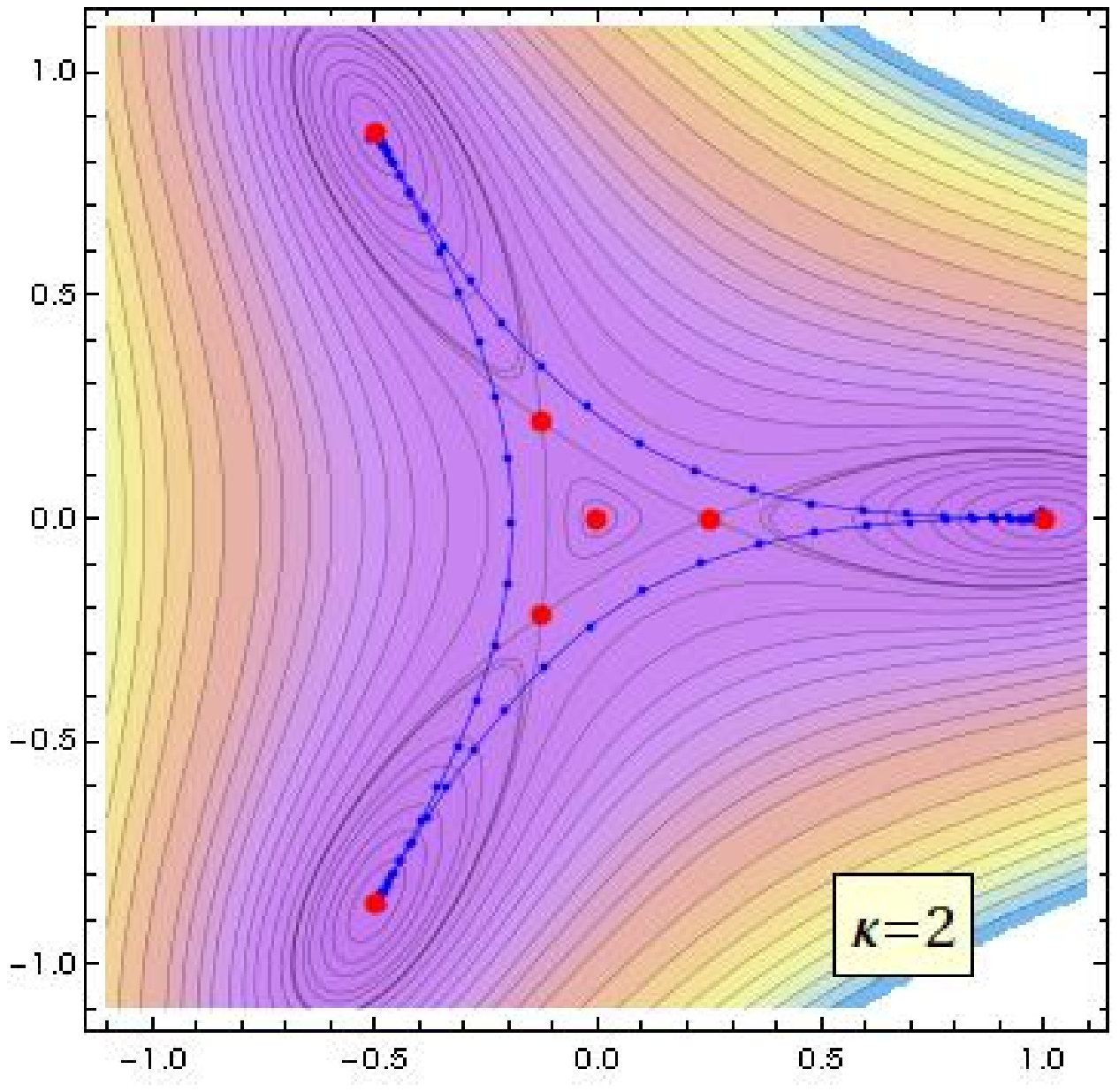}
\hfill
\includegraphics[width=7cm]{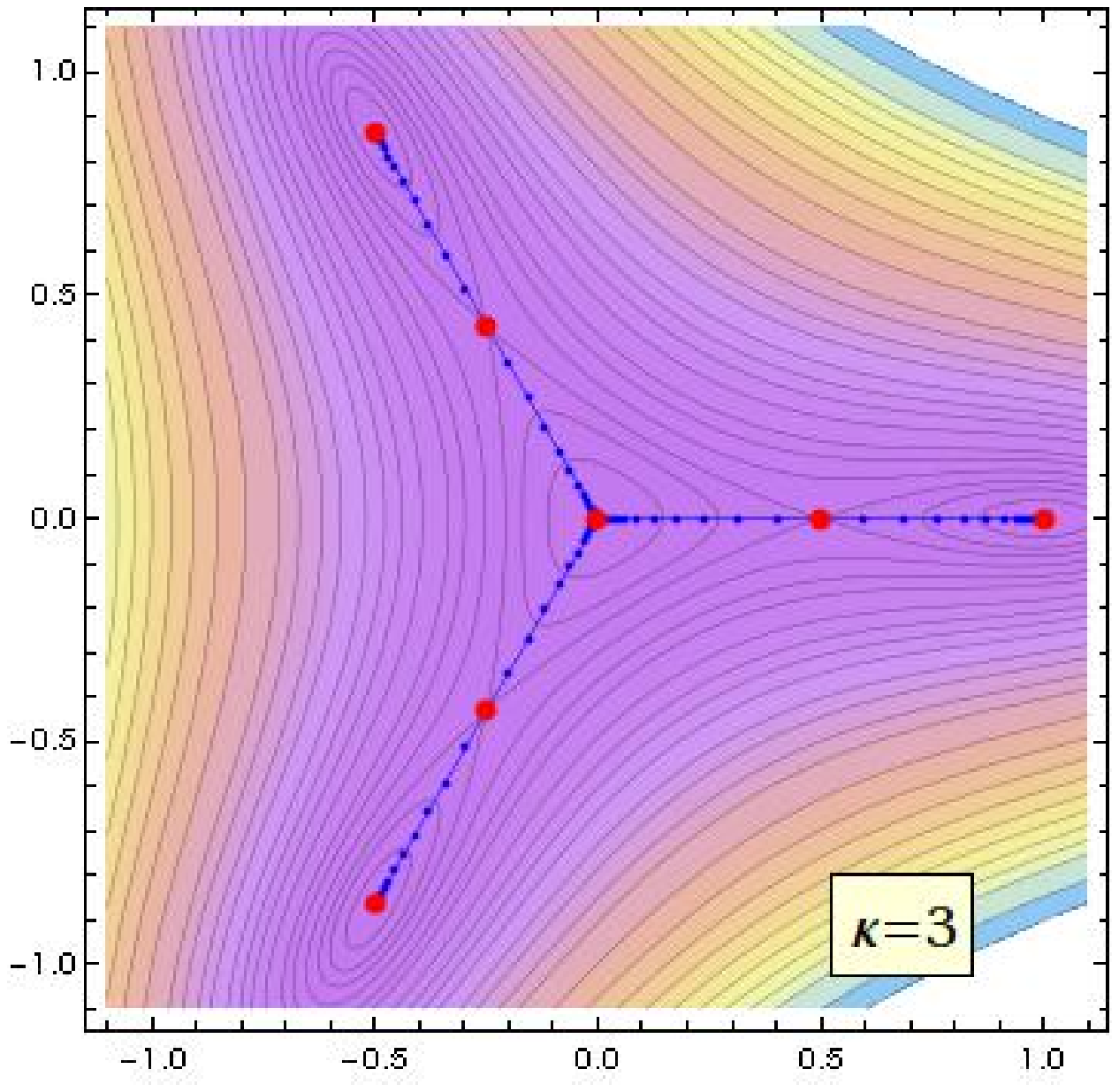}
}
\caption{Contour plots of $V(\phi_1{=}\phi_2{=}\phi_3)$, with critical points
and zero-energy kink trajectories.}
\label{fig:1}
\end{figure}

At the magical value of $\vk{=}{-}1$ the trajectories become straight,
and the solution analytic:
\begin{equation} \label{transverse}
\begin{aligned}
\phi_1(\tau)&\=
(\sfrac14{+}\im\sfrac{\sqrt{3}}4)+(-\sfrac34{+}\im\sfrac{\sqrt{3}}4)
\tanh(\sfrac{\tau{-}\tau_0}{2})\ ,\\[4pt]
\phi_2(\tau)&\=
-\sfrac12-\im\sfrac{\sqrt{3}}2\tanh(\sfrac{\tau{-}\tau_0}{2})\ ,\\[4pt]
\phi_3(\tau)&\=
(\sfrac14{-}\im\sfrac{\sqrt{3}}4)+(\sfrac34{+}\im\sfrac{\sqrt{3}}4)
\tanh(\sfrac{\tau{-}\tau_0}{2})\ .
\end{aligned}
\end{equation}

\noindent
{\bf Radial kinks at $\vk=3$. \ }
For this value of~$\vk$ the critial point at the origin is degenerate with
$(1,1,1)$ and its symmetry orbits. Therefore, we can connect any type~A 
critical point to the unique type~B point via `radial kinks', such as
\begin{equation} \label{radial}
\begin{aligned}
\phi_1(\tau) &\= \sfrac12 
\bigl(1+\tanh(\sfrac{\tau{-}\tau_0}{2\sqrt{3}})\bigr)\ ,\\[4pt]
\phi_2(\tau) &\= (-\sfrac14{+}\im\sfrac{\sqrt{3}}4)
\bigl(1+\tanh(\sfrac{\tau{-}\tau_0}{2\sqrt{3}})\bigr)\ ,\\[4pt]
\phi_3(\tau) &\= (-\sfrac14{-}\im\sfrac{\sqrt{3}}4)
\bigl(1+\tanh(\sfrac{\tau{-}\tau_0}{2\sqrt{3}})\bigr)\ ,\\[4pt]
\end{aligned}
\end{equation}
which connects
\begin{equation}
(0\,,0\,,0)\qquad\longrightarrow\qquad(1\,,\ep^{2\pi\im/3},\ep^{-2\pi\im/3})
\end{equation}
in a 3-symmetric fashion and is also marked in the lower right plot of Fig.~1.
It is the limiting case of the transverse kinks for $\vk\to+3$.
In the other limit, $\vk\to-3$, the particles move infinitely slowly on
the degenerate unit circle, $|\phi|=1$.

\noindent
{\bf Bounces at $\vk{<}{-}3$ and ${+}3{<}\vk{<}{+}5$. \ }
In the range $\vk\in(-\infty,-3)\cup(+3,+5)$ finite-action bounce solutions
must exist, in the form
\begin{equation} 
\phi_k(\tau)\=\ep^{2\pi\im(k-1)/3}\,f_\vk(\tau) \qquad\textrm{with}\quad
f_\vk(\pm\infty)=1 \and f_\vk(0)=\sfrac16\bigl(\vk{-}3+\sqrt{\vk^2{-}9}\bigr)
\ ,
\end{equation}
where $f_\vk(\tau)$ is a real function, so the trajectories are straight.
It is easy to find it numerically. Fig.~2 shows the trajectories for
$\vk=-4$ and $\vk=+4$.
\begin{figure}[ht]
\centerline{
\includegraphics[width=7cm]{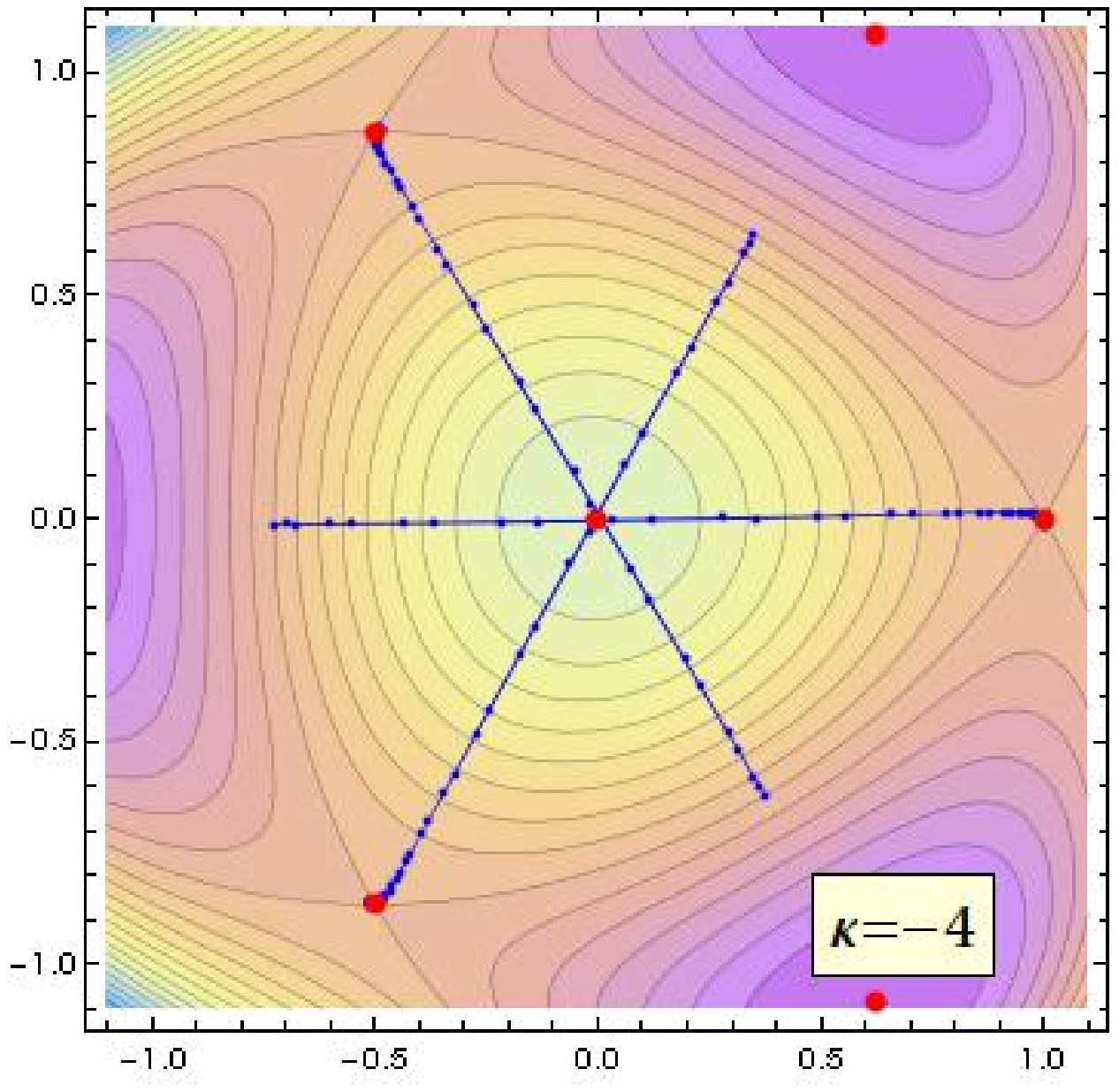}
\hfill
\includegraphics[width=7cm]{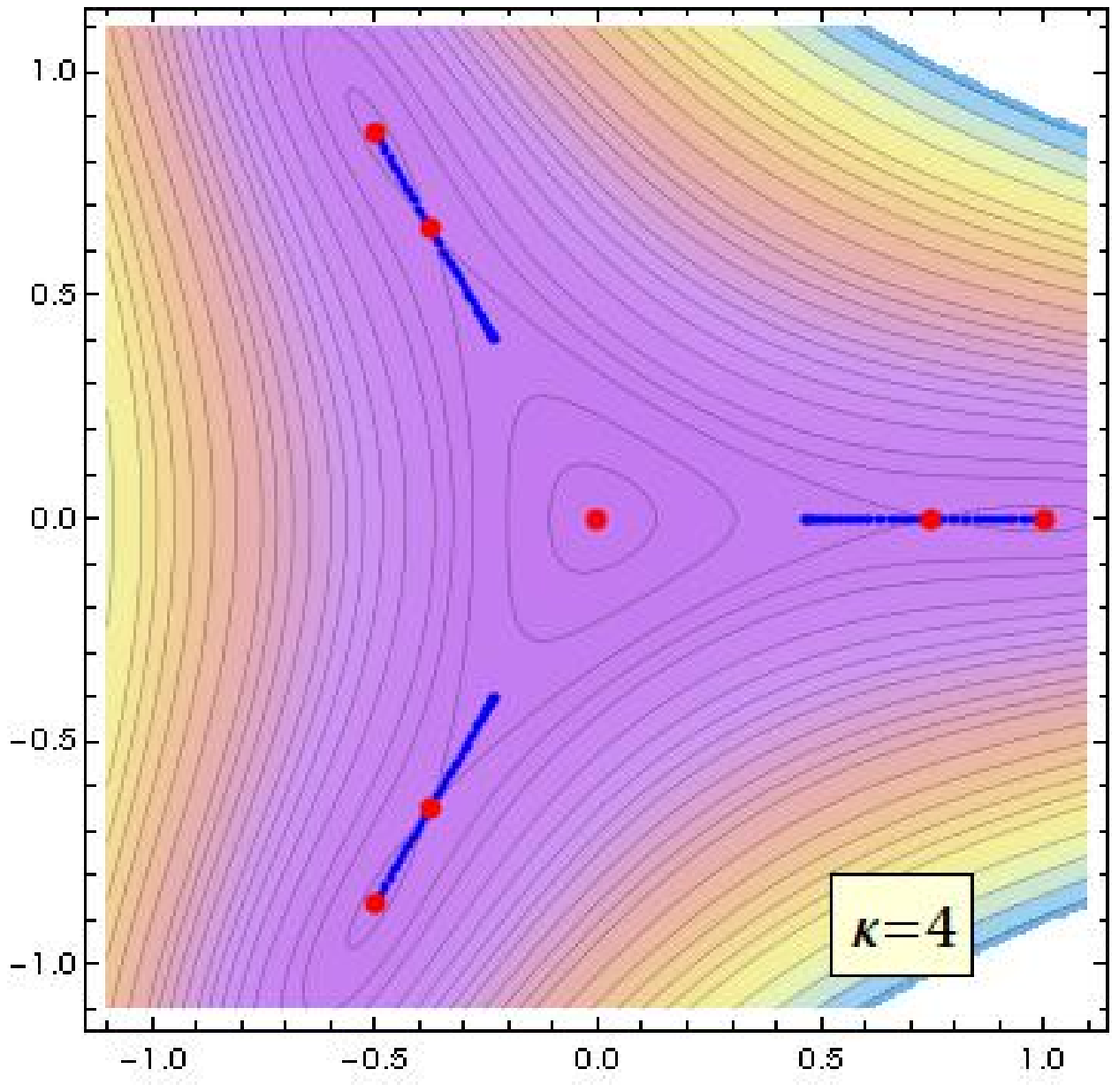}
}
\caption{Contour plots of $V(\phi_1{=}\phi_2{=}\phi_3)$, with critical points
and zero-energy bounce trajectories.}
\label{fig:2}
\end{figure}

\noindent
{\bf Radial bounce/kink at $\vk=-1{-}2\sqrt{3}$. \ }
If we put $\phi_1(\tau)=\phi_2(\tau)\equiv0$ at this $\vk$~value,
the remaining function is governed by the rotationally symmetric potential
\begin{equation}
V(0,0,\phi_3)\=2(2{+}\sqrt{3})\ -\ (1{+}\sqrt{3})|\phi_3|^2\ +\ |\phi_3|^4\ ,
\end{equation}
admitting the kink solution
\begin{equation} \label{bounce}
|\phi_3(\tau)| \= \textstyle{\sqrt{1{+}\sqrt{3}}}\,\tanh\bigl\{
\sqrt{\smash{\sfrac{1{+}\sqrt{3}}{6}}\phantom{{\big|}\!\!}}\,\tau\bigr\}
\qquad\textrm{while}\quad \phi_1(\tau)=\phi_2(\tau)\equiv0\ ,
\end{equation}
which interpolates between antipodal type~C critical points via point~B,
\begin{equation}
(0\,,0\,,-\ep^{\im\a}\textstyle{\sqrt{1{+}\sqrt{3}}}) 
\qquad\longrightarrow\qquad
(0\,,0\,,+\ep^{\im\a}\textstyle{\sqrt{1{+}\sqrt{3}}})\ .
\end{equation}

\section{Yang-Mills fields on $\R\times\,$Sp(2)/Sp(1)${\times}$U(1)}

\subsection{Explicit form of $X_a$ matrices}

The adjoint of Sp(2), restricted to Sp(1)${\times}$U(1), decomposes as
\begin{equation}
{\bf 10}\ (\textrm{of}\ \textrm{Sp}(2)) \= ({\bf 3}_0+{\bf 1}_0)_{\textrm{adj}}
+ {\bf 2}_{+1} + {\bf 2}_{-1} + {\bf 1}_{+2} + {\bf 1}_{-2}\ ,
\end{equation}
where the subscript denotes the U(1) charge.
Clearly, one has $q{=}2$ complex parameters.
As a convenient representation, 
let us take the fundamental $\Dcal={\bf 4}$ of Sp(2)\,$\subset$\,U(4).
Again, it turns out that $\chi_{\bf 4}/\chi_{\bf 10}=1/6$.

We choose the generators of the subgroup Sp(1)${\times}$U(1) of Sp(2)
in the form
\begin{equation}\label{5.1}
I_{7,8,9}\=\frac{\im}{2\sqrt{3}}
\begin{pmatrix}\s_{1,2,3}&{\bf 0}_2\\
{\bf 0}_2&{\bf 0}_2\end{pmatrix}
\und
I_{10}\=\frac{\im}{2\sqrt{3}}
\begin{pmatrix}{\bf 0}_2&{\bf 0}_2\\
{\bf 0}_2&\s_3\end{pmatrix}\ .
\end{equation}
Then solutions of the Sp(2)-invariance conditions (\ref{2.26}) are given
by matrices
\begin{equation}\label{5.2}
\begin{aligned}
X_{1}\=\frac{1}{2\sqrt{6}} {\small\begin{pmatrix}
0&0&0&\!\!-\vp\\
0&0&\!\!-\bar\vp&0\\
0&\,\vp\,&0&0\\
\bar\vp&0&0&0
\end{pmatrix}}\ ,\quad
X_{2}\=\frac{1}{2\sqrt{6}} {\small\begin{pmatrix}
0&0&0&\im\vp\\
0&0&\!\!-\im\bar\vp&0\\
0&\!\!-\im\vp&0&0\\
\im\bar\vp&0&0&0
\end{pmatrix}}\ ,\\[8pt]
X_{3}\=\frac{1}{2\sqrt{6}} {\small\begin{pmatrix}
0&0&\!\!-\bar\vp&0\\
0&0&0&\,\vp\\
\vp&0&0&0\\
0&\!\!-\bar\vp&0&0
\end{pmatrix}}\ ,\quad
X_{4}\=\frac{-1}{2\sqrt{6}} {\small\begin{pmatrix}
0&0&\,\im\bar\vp\;&0\\
0&0&0&\im\vp\\
\im\vp&0&0&0\\
0&\;\im\bar\vp\,&0&0
\end{pmatrix}}\ ,\\[8pt]
X_{5}\=\frac{1}{2\sqrt{3}} {\small\begin{pmatrix}
0&\ 0\ &0&0\\
0&0&0&0\\
0&0&0&\ \bar\chi\\
0&0&\!\!-\chi&0
\end{pmatrix}}\ ,\quad
X_{6}\=\frac{1}{2\sqrt{3}} {\small\begin{pmatrix}
\ 0\ &\ 0\ &\ 0\ &0\\
0&0&0&0\\
0&0&0&\im\bar\chi\\
0&0&\im\chi&0
\end{pmatrix}}\ ,
\end{aligned}
\end{equation}
where $\vp$ and $\chi$ are complex-valued functions of $\tau$.
Note that the generators $\{I_a\}$ of the group Sp(2) are obtained
from (\ref{5.2}) if one put $\vp=1=\chi$. The choice (\ref{5.1}) and
(\ref{5.2}) agrees with the standard form (\ref{3.2}), (\ref{3.5})
and (\ref{3.11})-(\ref{3.15}) of the nearly K\"ahler structure on
the manifold Sp(2)/Sp(1)${\times}$U(1).

\subsection{Equations of motion}

The equations of motion for Sp(2)-invariant gauge fields on
$\R{\times}$Sp(2)/Sp(1)${\times}$U(1) are obtained by plugging
(\ref{5.2}) into (\ref{3.19}) and (\ref{3.20}). After tedious
calculations we get
\begin{equation}\label{5.3}
\begin{array}{l}
6\,\ddot\vp \= 
(\vk{-}1)\,\vp\ -\ (\vk{+}3)\,\bar\vp\bar\chi\ +\ 
(3|\vp|^2+\ \,|\chi|^2)\,\vp\ ,
\\[4pt]
6\,\ddot\chi \=
(\vk{-}1)\,\chi\ -\ (\vk{+}3)\,\bar\vp^2\ \;+\ 
(2|\vp|^2+2|\chi|^2)\,\chi\ ,
\end{array}
\end{equation}
and
\begin{equation}\label{5.4}
\vp\dot{\bar\vp}-\dot\vp\bar\vp\=\chi\dot{\bar\chi}-\dot\chi\bar\chi
\end{equation}
Notice that these equations follow from (\ref{4.6}), (\ref{4.7}) after
identification
\begin{equation}\label{5.5}
\phi_1=\phi_2=:\vp\und \phi_3=:\chi\ .
\end{equation}
Furthermore, substituting (\ref{5.2}) into the action functional
(\ref{3.22}), we obtain the Lagrangian
\begin{equation}\label{5.6}
18\,{\cal L}\=12|\dot\vp|^2{+}6|\dot\chi|^2 -
(\vk{-}3) +
(\vk{-}1)\bigl(2|\vp|^2{+}|\chi|^2\bigr) -
(\vk{+}3)\bigl(\vp^2\chi{+}\bar\vp^2\bar\chi\bigr) +
3|\vp|^4{+}2|\vp\chi|^2{+}|\chi|^4 \ ,
\end{equation}
which also follows from (\ref{4.4}) after identification (\ref{5.5}).
The equations (\ref{5.3}) are the Euler-Lagrange equations for the
Lagrangian (\ref{5.6}),
\begin{equation}
12\,\ddot\vp \= \frac{\partial V}{\partial\bar\vp} \und
6\,\ddot\chi \= \frac{\partial V}{\partial\bar\chi}\ ,
\end{equation}
and the constraint~(\ref{5.4}) derives from the U(1) symmetry
\begin{equation} \label{u1}
\bigl(\,\vp\,,\chi\bigr)\ \mapsto\ \bigl(\,
\ep^{\im\de}\vp\,,\,\ep^{-2\im\de}\chi\bigr)
\end{equation}
of the potential
\begin{equation}
V \= -(\vk{-}3)\ +\
(\vk{-}1)\bigl(2|\vp|^2{+}|\chi|^2\bigr)\ -\
(\vk{+}3)\bigl(\vp^2\chi{+}\bar\vp^2\bar\chi\bigr)\ +\
3|\vp|^4 + 2|\vp\chi|^2 + |\chi|^4\ .
\end{equation}

\subsection{Some solutions}

Clearly, the solutions to (\ref{5.3}) and~(\ref{5.4}) form a subset of
the solutions to (\ref{4.6}) and~(\ref{4.7}), namely those where two
functions coincide. Since in all examples of the previous section this
can be arranged by applying a U(1)$\times$U(1) transformation~(\ref{u1u1}),
one gets $\vp(\tau)=\chi(\tau)$ equal to any of the functions appearing
on the right-hand sides of (\ref{transverse}) and~(\ref{radial}) or
depicted in Fig.~1, after dialling the corresponding $\vk$~value. In addition,
(\ref{bounce}) translates to a solution with $\vp\equiv0$ and a kink~$\chi$.

\subsection{Specialization to $S^6$ and flow equations}

By further identification
\begin{equation}\label{5.7}
\phi_1=\phi_2=\phi_3=:\phi
\end{equation}
we resolve the constraint equations (\ref{4.7}) 
and reduce (\ref{4.6}) to the equation
\begin{equation}\label{5.8}
6\,\ddot\phi \=(\vk{-}1)\,\phi\ -\ (\vk{+}3)\,\bar\phi^2\ +\ 4|\phi|^2\phi
\= \frac13 \frac{\partial V}{\partial\bar\phi}
\end{equation}
with
\begin{equation} \label{g2pot}
V \= -(\vk{-}3)\ +\ 3(\vk{-}1)\,|\phi|^2
\ -\ (\vk{+}3)\bigl(\phi^3{+}\bar\phi^3\bigr)\ +\ 6\,|\phi|^4\ .
\end{equation}
The U(1) symmetry~(\ref{u1}) is broken to the discrete 3-symmetry.
Clearly, the Lagrangian~(\ref{4.4}) maps to
\begin{equation}
18\,{\cal L}\=
18\,|\dot\phi|^2\ +\ V(\phi)\ ,
\end{equation}
which describes $G_2$-invariant gauge fields on $\R\times S^6$, where
$S^6=G_2/\textrm{SU}(3)$~\cite{HILP}.
All is consistent with the decomposition
\begin{equation}
{\bf 14}\ (\textrm{of}\ G_2) \= 
{\bf 8}_{\textrm{adj}} + {\bf 3} +\bar{\bf 3}\ (\textrm{of}\ \textrm{SU}(3))\ .
\end{equation}
Obviously, any function on the right-hand sides of 
(\ref{transverse}) and~(\ref{radial}) or shown in Fig.~1 is a zero-energy 
solution~$\phi(\tau)$, as was already noticed in~\cite{HILP}.
Vice versa, any solution of (\ref{5.8}) gives a special solution to the
equations (\ref{5.3}), (\ref{5.4}) and (\ref{4.6}), (\ref{4.7}).

Let us for a moment investigate the possibility of straight-trajectory
solutions $\phi(\tau)\in\C$ to~(\ref{5.8}).
With a 3-symmetry transformation, any such solution can
be brought into a form where either $\textrm{Re}\phi(\tau)=\textrm{const}$
or $\textrm{Im}\phi(\tau)=\textrm{const}$.
Then, the vanishing of the left-hand side of Re(\ref{5.8}) yields
two conditions on $\textrm{Re}\phi$ and~$\vk$, whose solutions
follow a Hamiltonian flow~\cite{HILP}:
\begin{equation} \label{imphi}
\begin{aligned}
\vk=-1 \and & \textrm{Re}\phi=-\sfrac12 
&\quad\Rightarrow\qquad&
\sqrt{3}\,\textrm{Im}\dot\phi=\sfrac34-(\textrm{Im}\phi)^2 
&\qquad\Leftrightarrow\qquad& 
\sqrt{3}\,\dot\phi=\im\,(\bar\phi^2-\phi)\ ,\\
\vk=-3 \and & \textrm{Re}\phi=0 
&\quad\Rightarrow\qquad& 
\sqrt{3}\,\textrm{Im}\dot\phi=1-(\textrm{Im}\phi)^2 
&\qquad\Leftrightarrow\qquad& 
\sqrt{3}\,\dot\phi=\sfrac{\phi}{|\phi|}\,(1-|\phi|^2)\ ,\\
\vk=-7 \and & \textrm{Re}\phi=1 
&\quad\Rightarrow\qquad&
\sqrt{3}\,\textrm{Im}\dot\phi=3-(\textrm{Im}\phi)^2 
&\qquad\Leftrightarrow\qquad&
\sqrt{3}\,\dot\phi=\im\,(\bar\phi^2+2\phi)\ .
\end{aligned}
\end{equation}
On the other hand, for Im$\ddot\phi=0$ one finds
\begin{equation} \label{nokink}
\textrm{any $\vk$} \and \textrm{Im}\phi=0 
\qquad\Rightarrow\qquad
6\,\textrm{Re}\ddot\phi\=
(\vk{-}1)\textrm{Re}\phi-(\vk{+}3)(\textrm{Re}\phi)^2+4(\textrm{Re}\phi)^3
\=\frac13 \frac{\partial V_{\R}}{\partial\textrm{Re}\phi}\ ,
\end{equation}
with
\begin{equation} \label{repot}
V_{\R}\= \bigl(\textrm{Re}\phi-1\bigr)^2\,\bigl(
6(\textrm{Re}\phi)^2-(\vk{-}3)(2\textrm{Re}\phi+1)\bigr)\ .
\end{equation}
This includes the gradient-flow situations~\cite{HILP}
\begin{equation} \label{rephi}
\begin{aligned}
\vk=+3 \and & \textrm{Im}\phi=0 
&\quad\Rightarrow\qquad&
\sqrt{3}\,\textrm{Re}\dot\phi=(\textrm{Re}\phi)^2-\textrm{Re}\phi
&\qquad\Leftrightarrow\qquad&
\sqrt{3}\,\dot\phi=\bar\phi^2-\phi\ ,\\
\vk=+9 \and & \textrm{Im}\phi=0
&\quad\Rightarrow\qquad&
\sqrt{3}\,\textrm{Re}\dot\phi=(\textrm{Re}\phi)^2-2\,\textrm{Re}\phi
&\qquad\Leftrightarrow\qquad&
\sqrt{3}\,\dot\phi=\bar\phi^2-2\phi\ .
\end{aligned}
\end{equation}
All kink solutions to (\ref{imphi}) and~(\ref{rephi}) 
were given in~\cite{HILP}.
They have zero energy and thus finite action only for $\vk=-3$, $-1$ and~$+3$.
The latter two cases are also displayed in (\ref{transverse}) and
(\ref{radial}), respectively.
In addition, for $\vk{<}{-}3$ and ${+}3{<}\vk{<}{+}5$ one can also 
numerically construct finite-action bounce solutions to~(\ref{nokink}).

\medskip

\noindent
{\bf Remark.} Note that a nearly K\"ahler structure exists also on the
space $S^3\times S^3$. However, we do not consider the Yang-Mills equations
on $\R\times S^3\times S^3$ since this was already done in~\cite{IL}.

\section{Instanton-anti-instanton chains and dyons}

If we replace $\R\times G/H$ with $S^1\times G/H$, the time interval will be 
of finite length, namely the circle circumference~$L$, 
and we are after solutions periodic in~$\tau$.
In this case, the action is always finite, and the $E{=}0$ requirement gets
replaced by $\phi_i(\tau{+}L)=\phi_i(\tau)$. The physical interpretation of
such configurations is one of instanton-anti-instanton chains.

\subsection{Periodic solutions}

As the simplest case we take $G/H=G_2/\textrm{SU}(3)$ 
and consider the magical $\vk$~values which admit analytic solutions
for~$\phi(\tau)\in\C$.
Switching from $\tau\in\R$ to $\tau\in S^1$, 
we must impose the periodicity conditions
\begin{equation}\label{6.4}
\phi(\tau{+}L)\=\phi(\tau)
\end{equation}
not on the flow equations (\ref{imphi}) and~(\ref{rephi})
but on the corresponding second-order equations,
\begin{equation} \label{sphaleroneom}
\begin{aligned}
\vk=-1 \and \textrm{Re}\phi=-\sfrac12 \,
&\qquad\Rightarrow\qquad
\sfrac32\,\textrm{Im}\ddot\phi\=
\textrm{Im}\phi\,(\textrm{Im}\phi^2-\sfrac34)\ ,\\
\vk=-3 \and \textrm{Re}\phi=0 \quad
&\qquad\Rightarrow\qquad 
\sfrac32\,\textrm{Im}\ddot\phi\=
\textrm{Im}\phi\,(\textrm{Im}\phi^2-1)\ ,\\
\vk=-7 \and \textrm{Re}\phi=1 \quad
&\qquad\Rightarrow\qquad
\sfrac32\,\textrm{Im}\ddot\phi\=
\textrm{Im}\phi\,(\textrm{Im}\phi^2-3)\ ,\\
\vk=+3 \and \textrm{Im}\phi=0 \quad
&\qquad\Rightarrow\qquad
\sfrac32\,\textrm{Re}\ddot\phi\=
\textrm{Re}\phi\,(\textrm{Re}\phi-\sfrac12)\,(\textrm{Re}\phi-1)\ ,\\
\vk=+9 \and \textrm{Im}\phi=0 \quad
&\qquad\Rightarrow\qquad
\sfrac32\,\textrm{Re}\ddot\phi\=
\textrm{Re}\phi\,(\textrm{Re}\phi-1)\,(\textrm{Re}\phi-2)\ .
\end{aligned}
\end{equation}
At finite~$L$, we obtain a different kind of solution (sphalerons), namely
\begin{equation}\label{6.5}
\begin{aligned}
\phi(\tau)&\=\beta\pm\im\sqrt{3}\,\gamma\,k\,b(k)\;\sn[b(k)\gamma\tau; k]
\quad&\textrm{with}\quad
&(\vk;\beta,\gamma)=(-1;-\sfrac12,1),\ (-3;0,\sfrac2{\sqrt{3}}),\ (-7;1,2)\ ,\\
\phi(\tau)&\=\beta\pm\;\sqrt{3}\,\gamma\,k\,b(k)\;\sn[b(k)\gamma\tau; k]
\quad&\textrm{with}\quad
&(\vk;\beta,\gamma)=(+3;\sfrac12,\sfrac1{\sqrt{3}}),\ (+9;1,\sfrac2{\sqrt{3}})
\ .
\end{aligned}
\end{equation}
Here $b(k)=(2{+}2k^2)^{-1/2}$ and $0\le k\le 1$.
Since the Jacobi elliptic function $\sn[u;k]$ has a period of $4{K}(k)$
(see Appendix~B), the condition (\ref{6.4}) is satisfied if
\begin{equation}\label{6.6}
\gamma\,b(k)\,L\=4{K}(k)\,n \qquad\textrm{for}\quad n\in\N\ ,
\end{equation}
which fixes $k=k(L,n)$ so that 
$\phi(\tau;k(L,n))=:\phi^{(n)}(\tau)$.
Solutions (\ref{6.5}) exist if $L\ge 2\pi\sqrt{2}\,n$~\cite{group8}.

By virtue of the periodic boundary conditions (\ref{6.4}),
the topological charge of the sphaleron $\phi^{(n)}$ is zero.
In fact, the configuration is interpreted as a chain
of $n$ kinks and $n$ antikinks, alternating and equally spaced
around the circle~\cite{MS, group8}. 
Interpreted as a static configuration on $S^1\times G/H$,
the energy of the sphaleron is
\begin{equation}\label{6.8}
{\cal E}\=\int\limits_0^L\!\diff\tau\left\{|\dot\phi|^2+V(\phi)\right\}
\end{equation}
and e.g. for the case of~$\vk=-3$ in (\ref{6.5}) we obtain
\begin{equation}\label{6.9}
{\cal E}[\phi^{(n)}]\=\frac{2n}{3\sqrt{2}}\bigl[8(1{+}k^2)\,E(k)\ -\
(1{-}k^2)(5{+}3k^2)\,{K}(k)\bigr]\ ,
\end{equation}
where ${K}(k)$ and $E(k)$ are the complete elliptic integrals of
the first and second kind, respectively~\cite{group8}.

The non-BPS solutions (\ref{6.5}) can be embedded into
the other cosets $G/H$, where they are special solutions, with
$\vp=\chi$ or $\phi_1=\phi_2=\phi_3$, respectively. Their degeneracy may
be lifted by applying a symmetry transformation (\ref{u1}) or~(\ref{u1u1}),
respectively.
Substituting our non-BPS solutions into (\ref{4.3}) or (\ref{5.2})
and then into~(\ref{2.25}), 
we obtain a finite-action Yang-Mills configuration which is
interpreted as a chain of $n$ instanton-anti-instanton pairs sitting on
$S^1\times G/H$ with six-dimensional nearly K\"ahler coset space $G/H$.
Away from the magical $\vk$~values, such chains are to be found numerically.

\subsection{Dyonic solutions}

Let us finally change the signature of the metric on
$\R\times G/H$ from Euclidean to Lorentzian by choosing on $\R$ a
coordinate $t=-\im\tau$  so that $\tilde e^0 = \diff t= -\im\diff\tau$.
Then as metric on $\R\times G/H$ we have
\begin{equation}\label{6.10}
\diff s^2 \= -(\tilde e^0)^2 +\de_{ab}e^ae^b\ .
\end{equation}
The $G$-invariant solutions (\ref{4.3}) and~(\ref{5.2}) for the matrices
$X_a$ are not changed. After substituting them into the Yang-Mills
equations on $\R\times G/H$, we arrive at the same second-order differential
equations as in the Euclidean case, except for the replacement
\begin{equation}\label{6.11}
\ddot\phi_i \qquad\longrightarrow\qquad -\sfrac{\diff^2\phi_i}{\diff t^2}\ .
\end{equation}
In particular, this implies a sign change of the left-hand side relative to 
the right-hand side in (\ref{4.6}), (\ref{5.3}) and (\ref{5.8}).
Thus, in the Lagrangians we effectively have a sign flip of the potential~$V$,
so that the analog Newtonian dynamics for $(\phi_i(t))$ is based on $+V$.

Let us again for simplicity look at the case of $G/H=G_2/\textrm{SU}(3)$.
Although the Lorentzian variant of (\ref{5.8}),
\begin{equation} \label{lorentz}
6\,\frac{\diff^2\phi}{\diff t^2} \=
-(\vk{-}1)\,\phi\ +\ (\vk{+}3)\,\bar\phi^2\ -\ 4|\phi|^2\phi
\= -\frac13 \frac{\partial V}{\partial\bar\phi}
\end{equation}
with $V$ from~(\ref{g2pot}), does not follow
from first-order equations for any of the magical values 
$\vk=-1$, $-3$, $-7$, $+3$ or $+9$, 
it can still be explicitly integrated in those cases,
\begin{equation}\label{6.12}
\begin{aligned}
\phi(t)&\=\beta\pm\im\sqrt{\sfrac32}\,\gamma\,
\cosh^{-1}\!\sfrac{\gamma\,t}{\sqrt{2}} \qquad&\textrm{with}\quad
&(\vk;\beta,\gamma)=(-1;-\sfrac12,1),\ (-3;0,\sfrac2{\sqrt{3}}),\ (-7;1,2)\ ,\\
\phi(t)&\=\beta\pm\sqrt{\sfrac32}\,\gamma\,
\cosh^{-1}\!\sfrac{\gamma\,t}{\sqrt{2}} \qquad&\textrm{with}\quad
&(\vk;\beta,\gamma)=(+3;\sfrac12,\sfrac1{\sqrt{3}}),\ (+9;1,\sfrac2{\sqrt{3}})
\ .
\end{aligned}
\end{equation}
The 3-symmetry action maps these solutions to rotated ones.
Any such configuration is a bounce in our double-well-type potential,
which most of the time hovers around a saddle point.
For other values of~$\vk$, such bounce solutions may be found numerically.

Inserting (\ref{6.12}) into the gauge potential, we arrive at
dyon-type configurations with smooth nonvanishing `electric' and `magnetic'
field strength $\Fcal_{0a}$ and $\Fcal_{ab}$, respectively. The total energy
\begin{equation}\label{6.13}
-\tr\,(2\Fcal_{0a}\Fcal_{0a}+\Fcal_{ab}\Fcal_{ab})\times{\rm Vol}(G/H)
\end{equation}
for these configurations is finite, but their action diverges
unless $\phi(\pm\infty)=\ep^{2\pi\im k/3}$. These are saddle points
for $\vk<-3$ and $\vk>+5$. Thus, for $|\vk{-}1|>4$ the potential~(\ref{g2pot})
admits pairs~$\phi_\pm(t)$ of finite-action dyons, with
\begin{equation}
\phi_\pm(\pm\infty)\=1 \und 
\phi_\pm(0)\=
\sfrac16\bigl(\vk{-}3\pm\sqrt{\vk^2{-}9}\bigr) 
\qquad\textrm{for}\quad \vk>+5
\end{equation}
and a more complex behavior for $\vk<-3$.
The $\vk{=}{-}7$ and $\vk{=}{+}9$ straight-line solutions in (\ref{6.12}) 
are among these. Numerical trajectories for some intermediate values are
shown in the plots of Figure~3.
\begin{figure}[ht]
\centerline{
\includegraphics[width=7cm]{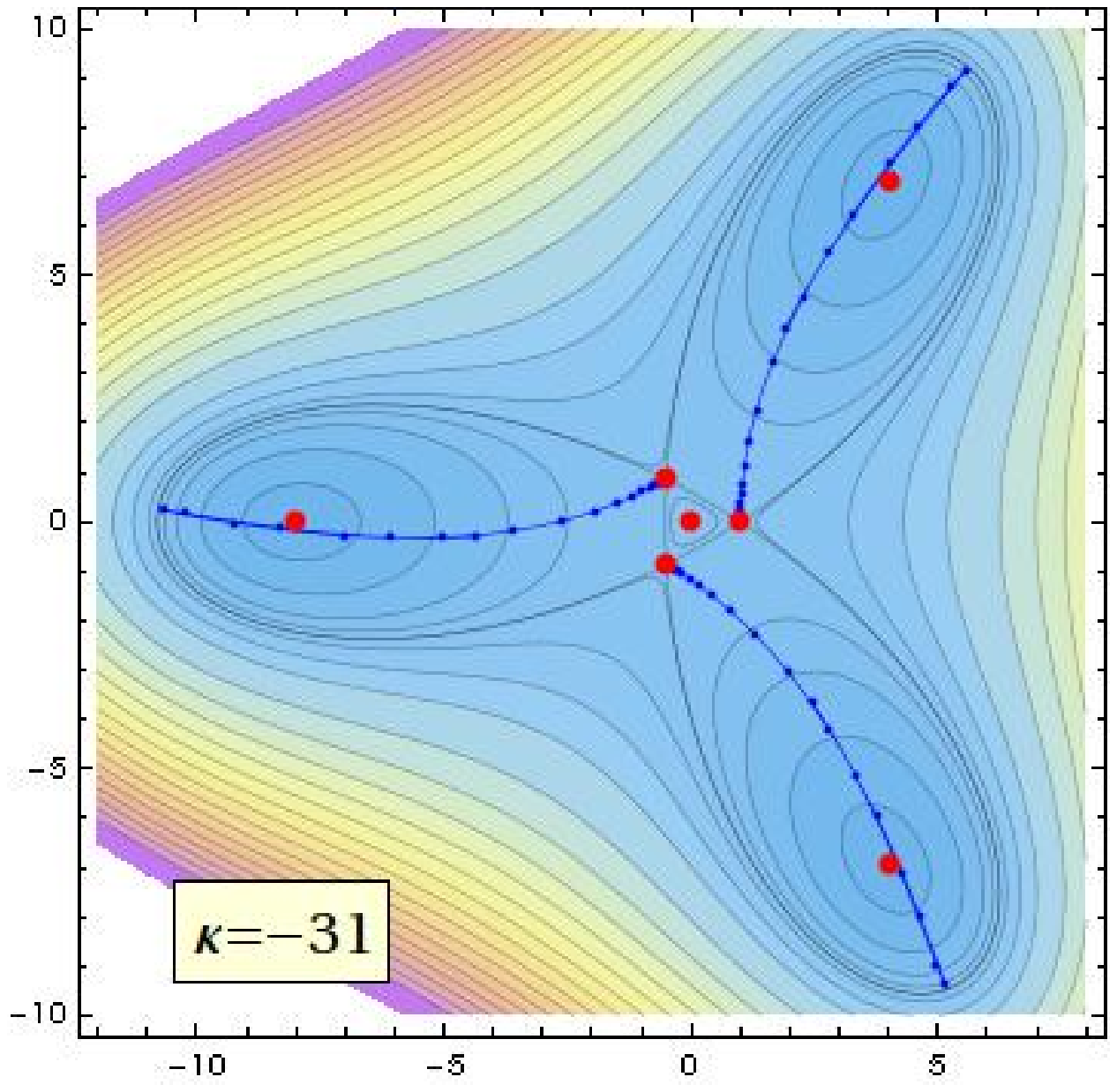}
\hfill
\includegraphics[width=7cm]{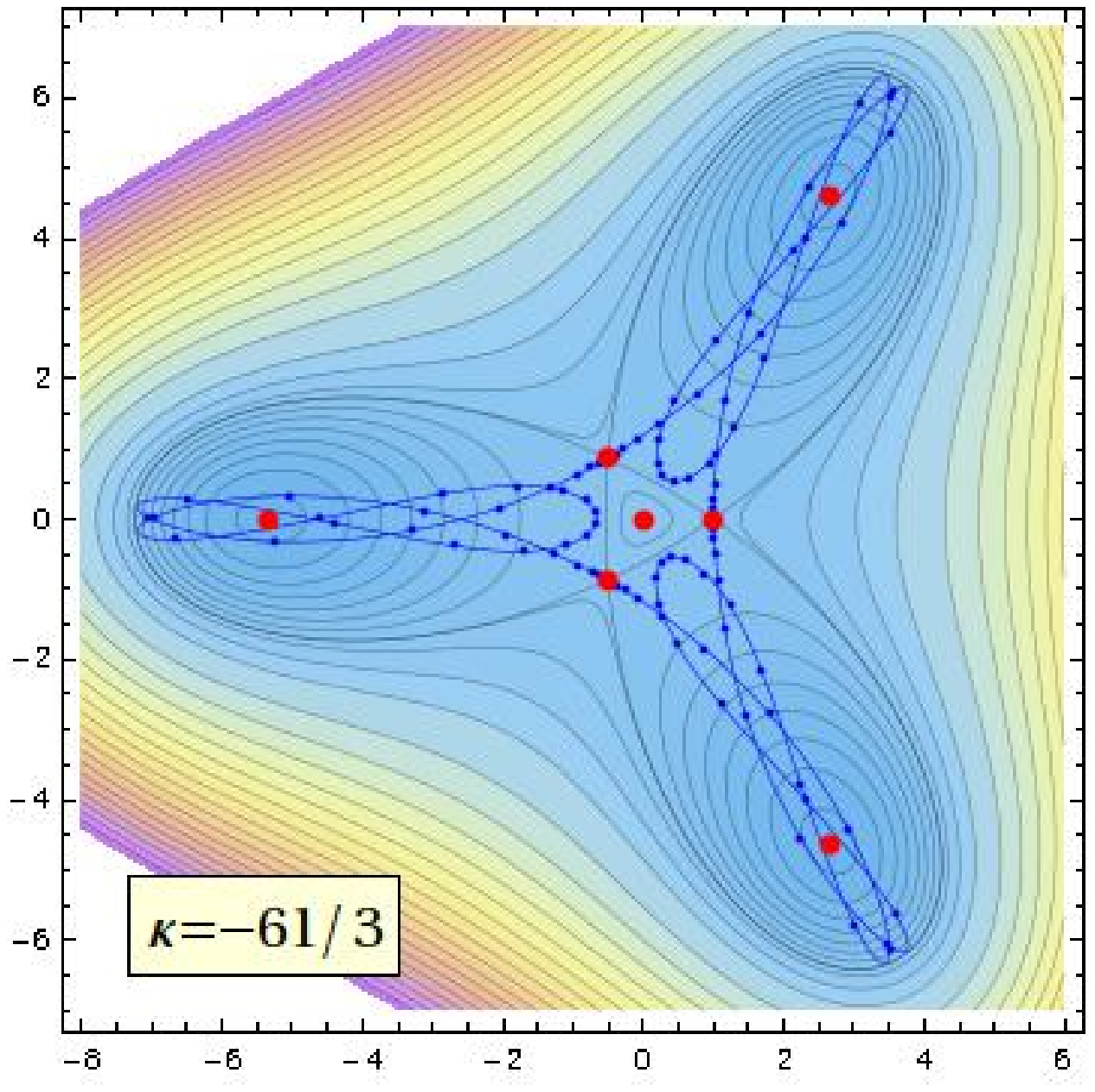}
}
\bigskip
\centerline{
\includegraphics[width=7cm]{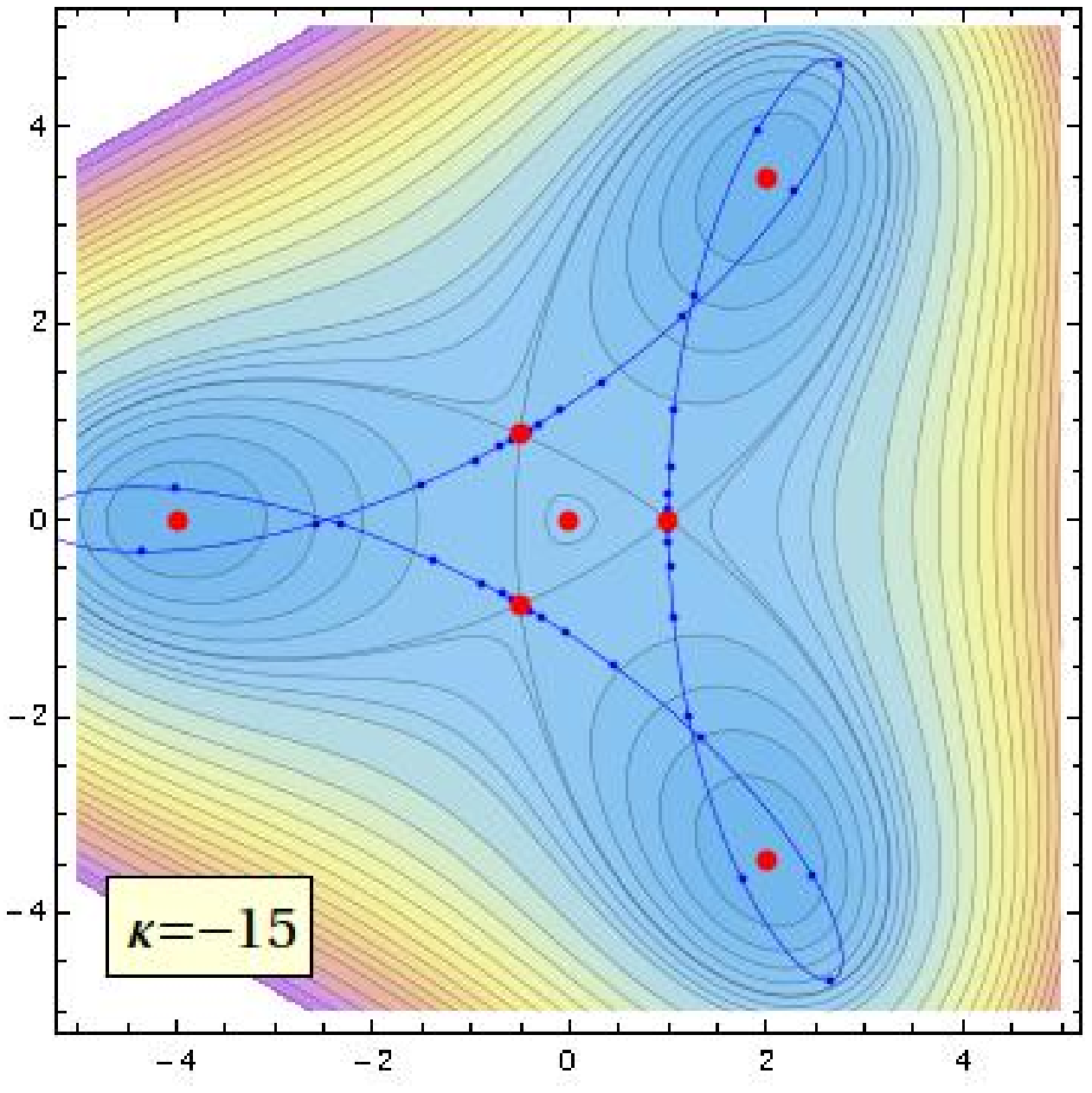}
\hfill
\includegraphics[width=7cm]{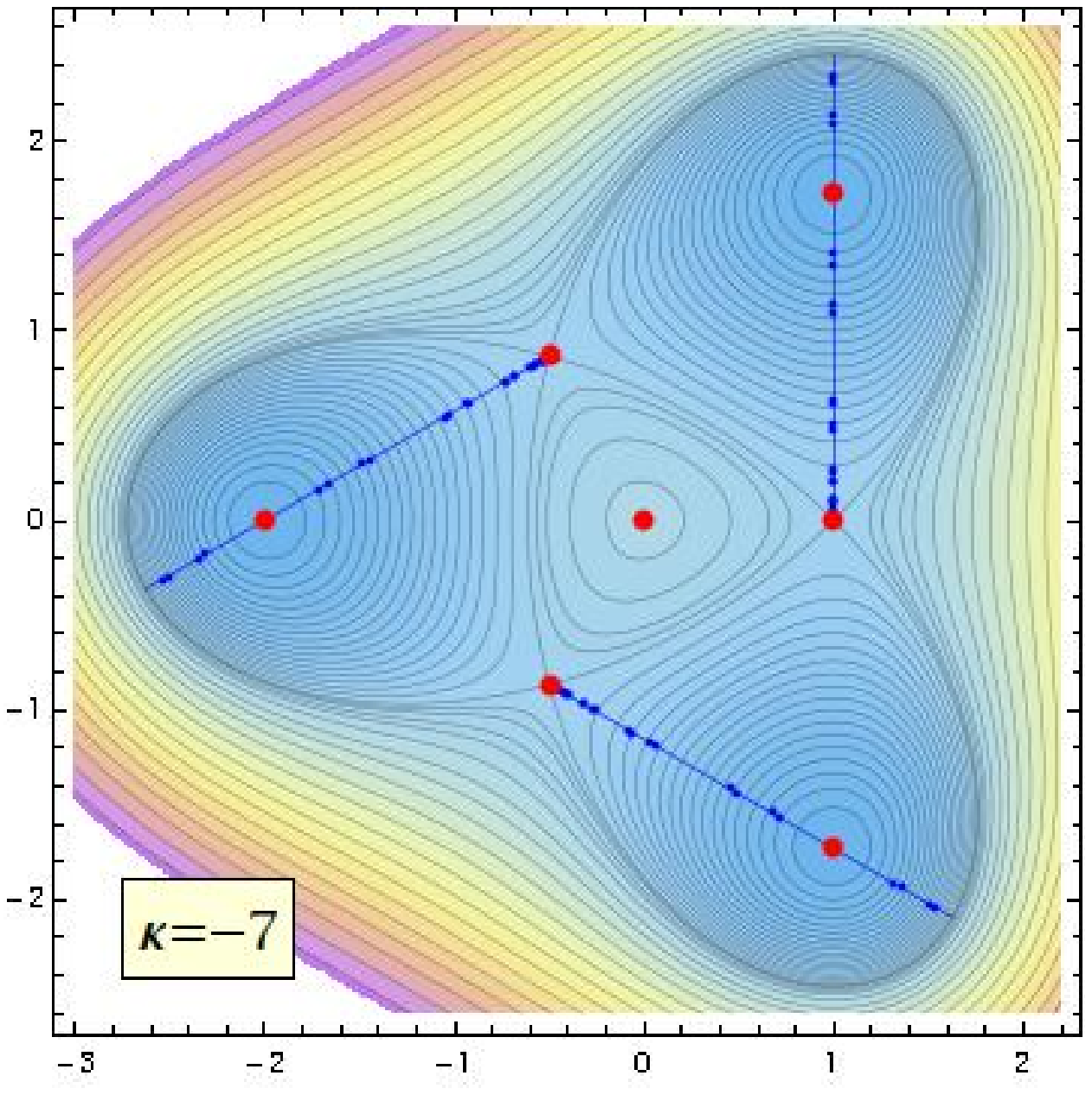}
}
\bigskip
\centerline{
\includegraphics[width=7cm]{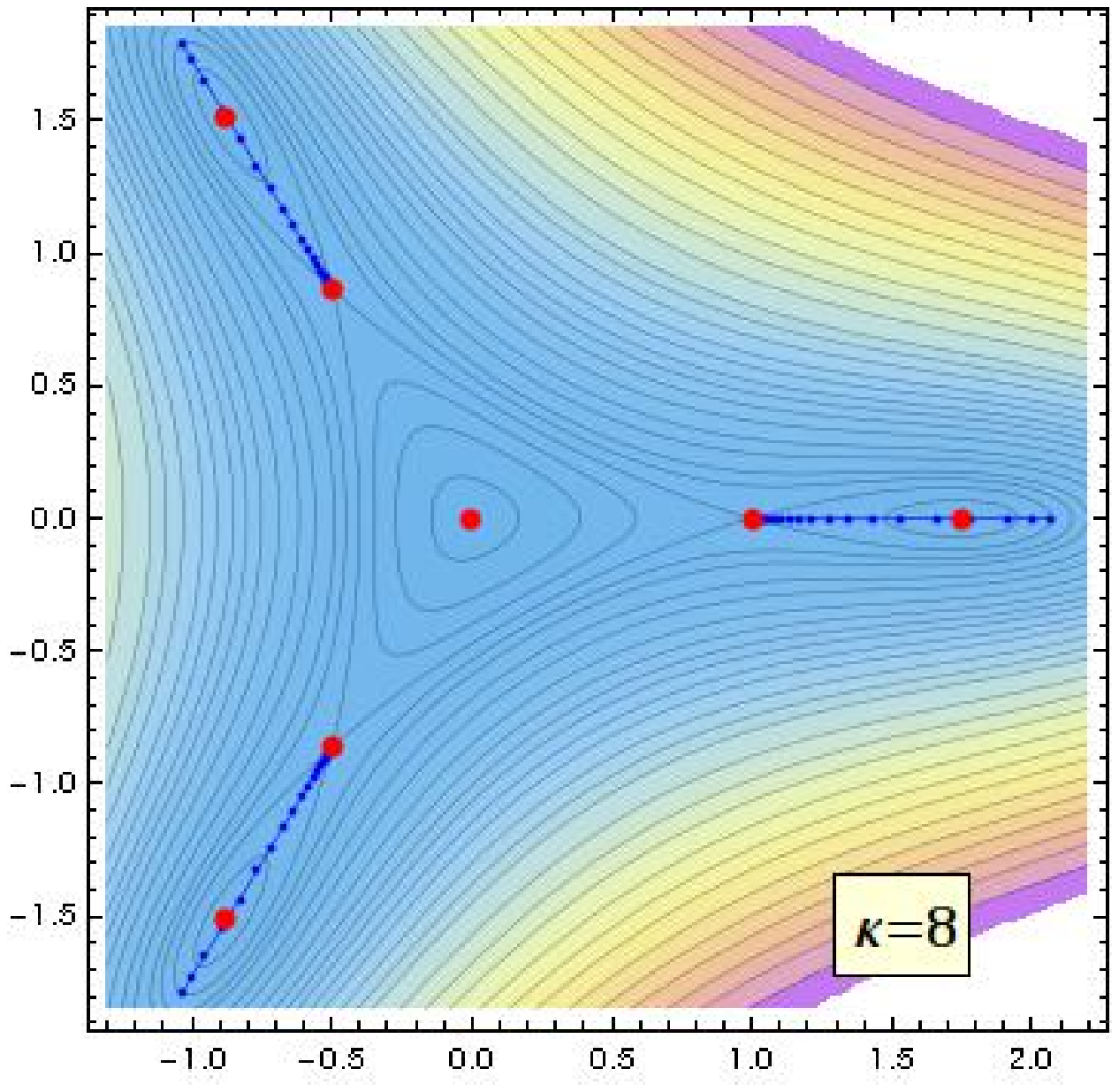}
\hfill
\includegraphics[width=7cm]{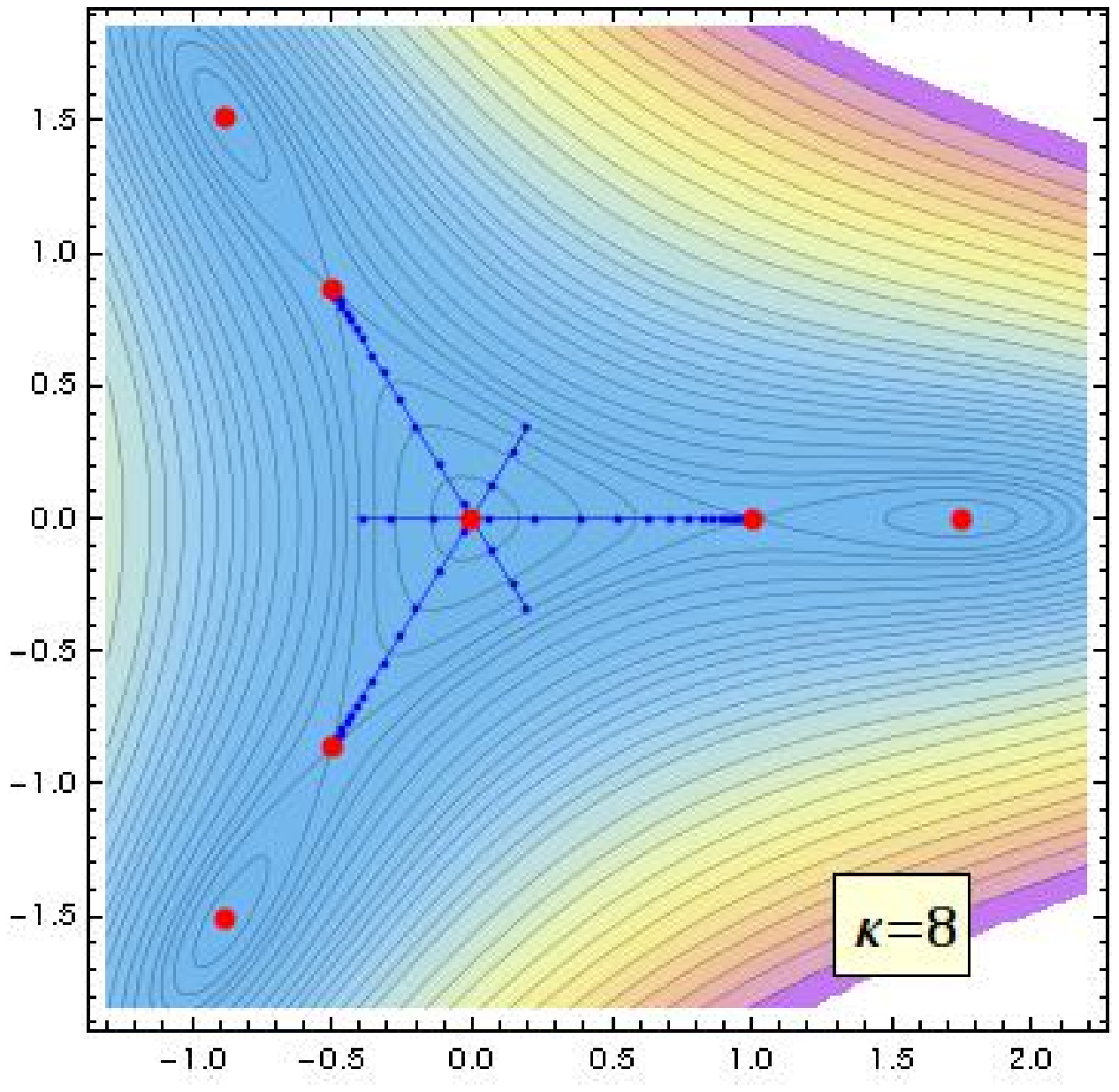}
}
\caption{Contour plots of $V(\phi_1{=}\phi_2{=}\phi_3)$, with critical points
and finite-action dyon trajectories.}
\label{fig:3}
\end{figure}

\bigskip

\noindent
{\bf Acknowledgements}

\medskip

\noindent
The authors are grateful to Alexander Popov for fruitful discussions and 
useful comments. O.L.\ thanks N.~Dragon for remarks on the critical points.
This work was supported in part by the Deutsche Forschungsgemeinschaft (DFG),
by the Russian Foundation for Basic Research (grant RFBR 09-02-91347)
and by the Heisenberg-Landau program.

\newpage

\appendix

\noindent {\bf \large Appendix~A. Zero-energy critical points}

\renewcommand{\thesection}{\Alph{section}.}
\setcounter{subsection}{0} \setcounter{equation}{0}
\renewcommand{\thesubsection}{\Alph{subsection}}
\renewcommand{\theequation}{\rm{A}.\arabic{equation}}

\noindent
Here, we prove that the table in Subsection 4.3 lists all zero-energy
critical points~$(\ph_1,\ph_2,\ph_3)$ of the potential~(\ref{V3}), 
modulo permutations of the $\ph_i$ and actions of the U(1)${\times}$U(1) 
symmetry~(\ref{u1u1}). 

With the help of this symmetry, we can remove the phases of 
$\ph_1$ and~$\ph_2$. Since it was already argued that extremality implies
$\sum_i\arg\ph_i=0$ or $\pi$, also $\ph_3$ must be real.
Hence, we may take
\begin{equation}
\ph_1\,,\ph_2\ \in\ \R_+ \and \ph_3\ \in \R
\end{equation}
and investigate the solution space of $\diff{V}{=}0{=}V$, i.e.
\begin{eqnarray}
&&\!\!\!\!\!\!\!\!
(\vk{-}1)\,\ph_i-(\vk{+}3)\ph_j\ph_k+(2\ph_i^2+\ph_j^2+\ph_k^2)\,\ph_i
\=0 \quad\textrm{for}\quad i\neq j\neq k\in\{1,2,3\} 
\und \label{dV=0} \\[4pt] \label{V=0}
&&\!\!\!\!\!\!\!\!
(\vk{-}1){\textstyle\sum_i}\ph_i^2-2(\vk{+}3)\,\ph_1\ph_2\ph_3
+{\textstyle\sum_i}\ph_i^4+{\textstyle\sum_{i<j}}\ph_i^2\ph_j^2
\=\vk{-}3\ .
\end{eqnarray}

Let us first look at the exceptional cases where one of the $\ph_i$ vanishes.
{}From~(\ref{dV=0}) it follows that $\ph_i=0$ implies $\ph_j\ph_k=0$.
The trivial solution is
\begin{equation}
\ph_1=\ph_2=\ph_3=0 
\qquad\buildrel(\ref{V=0})\over\Rightarrow\qquad \vk=3
\end{equation}
and is labelled as type~B in the table. Generically, however, we have
\begin{equation}
\ph_1=\ph_2=0 \and \ph_3\neq0 
\qquad\buildrel(\ref{dV=0})\over\Rightarrow\qquad
\vk{-}1+2\,\ph_3^2=0
\qquad\buildrel(\ref{V=0})\over\Rightarrow\qquad
\vk=-1\pm2\sqrt{3}
\end{equation}
and reproduce type~C in the table.\footnote{
Only one of the two values for $\vk$ leads to a real~$\ph_3$.}

It remains to study the situation where all $\ph_i$ are nonzero.
Multiplying (\ref{dV=0}) with $\ph_i$ and taking the difference of any two
of the resulting three equations, we obtain the three conditions
\begin{equation} \label{cond1}
\bigl(\vk{-}1+2\ph_i^2+2\ph_j^2+\ph_k^2\bigr)\,\bigl(\ph_i^2-\ph_j^2\bigr)
\=0\ .
\end{equation}
Likewise, multiplying (\ref{dV=0}) with $\ph_j\ph_k$ and taking the difference
of any two of those three equations, we find three more conditions,
\begin{equation} \label{cond2}
\bigl((\vk{+}3)\,\ph_k^2+\ph_1\ph_2\ph_3\bigr)\,\bigl(\ph_i^2-\ph_j^2\bigr)
\=0\ .
\end{equation}
A little thought reveals that there are only two options. The first one is
\begin{equation}
\ph_1^2=\ph_2^2=\ph_3^2 \qquad\Rightarrow\qquad
\ph_1=\ph_2=\pm\ph_3\ =:\ \ph\in\R_+\ .
\end{equation}
The potential on this subspace becomes
\begin{equation}
V(\ph,\ph,\pm\ph)\=
\bigl( 6\,\ph^2\mp(\vk{-}3)(2\ph-1) \bigr)\,\bigl(\ph\mp1\bigr)^2\ ,
\end{equation}
and its critical zeros on the positive real axis are
\begin{equation}
(\ph_1,\ph_2,\ph_3;\ \vk)\=
(+1,+1,+1;\ \textrm{any})\quad\textrm{and}\quad(+1,+1,-1;\ -3)
\end{equation}
for the two sign choices, respectively. 
We have recovered types A and~A' of our table.

The second option for fulfilling (\ref{cond1}) and~(\ref{cond2}) is,
modulo permutation,
\begin{equation}
\ph_1^2=\ph_2^2\neq\ph_3^2 \qquad\Rightarrow\qquad
\ph_1=\ph_2\ =:\ \vph\in\R_+ \and \ph_3\ =:\ \ch\in\R\ , 
\end{equation}
with the simultaneous requirements
\begin{equation}
\vk{-}1+3\vph^2+2\ch^2\=0 \und \vk{+}3+\ch\=0
\end{equation}
from (\ref{cond1}) and (\ref{cond2}), respectively. The solution 
\begin{equation}
\vph\=\sqrt{-\sfrac23\vk^2-\sfrac{13}3\vk-\sfrac{17}3} \und \ch\=-\vk-3
\end{equation}
restricts $-13{-}\sqrt{33}<4\vk<-13{+}\sqrt{33}$, but one finds that
\begin{equation}
V(\vph,\vph,\ch)\= -\sfrac13\,(\vk{+}1)\,(\vk{+}4)^3\ ,
\end{equation}
which leaves only
\begin{equation}
\vk=-4 \qquad\Rightarrow\qquad \vph=\ch=1\ ,
\end{equation}
falling back to type~A.
Thus, the list of critical zeros presented in Subsection~4.3 is exhaustive.

\noindent {\bf \large Appendix~B. Jacobi elliptic functions}

\renewcommand{\thesection}{\Alph{section}.}
\setcounter{subsection}{0} \setcounter{equation}{0}
\renewcommand{\thesubsection}{\Alph{subsection}}
\renewcommand{\theequation}{\rm{B}.\arabic{equation}}

\noindent
The Jacobi elliptic functions arise from the inversion of the
elliptic integral of the first kind,
\begin{equation}
u\=F(\xi,k)\= \int\limits_{0}^{\xi}\frac{\diff x}{\sqrt{1-k^{2}\sin x}}\ ,
\qquad 0\le k^{2}<1\ ,
\end{equation}
where $k=\mathrm{mod}\, u$ is the elliptic modulus 
and $\xi=\am(u,k)=\am(u)$ is the Jacobi amplitude, giving
\begin{equation}
\xi\=F^{-1}(u,k)\=\am(u,k)\ .
\end{equation}
Then the three basic functions $\sn$, $\cn$ and $\dn$ are defined by
\begin{align}
\sn[u;k]&\=\sin(\am(u,k))\=\sin\xi\ ,\\
\cn[u;k]&\=\cos(\am(u,k))\=\cos\xi\ ,\\
\dn[u;k]^2&\=1-k^{2}\sin^{2}(\am(u,k))\=1-k^{2}\sin^{2}\xi\ .
\end{align}
These functions are periodic in $K(k)$ and $\tilde{K}(k)$,
\begin{align}
\sn[u{+}2mK{+}2n\im\tilde{K};k]&\=(-1)^{m}\sn[u;k]\ ,\\
\cn[u{+}2mK{+}2n\im\tilde{K};k]&\=(-1)^{m+n}\cn[u;k]\ ,\\
\dn[u{+}2mK{+}2n\im\tilde{K};k]&\=(-1)^{n}\dn[u;k]\ ,
\end{align}
where $K(k)$ is the complete elliptic integral of the first kind, 
\begin{equation}
K(k)\ :=\ F(\sfrac{\pi}{2},k) \und 
\tilde{K}(k)\ :=\ K(\sqrt{1{-}\smash{k^2}}) 
\= F(\sfrac{\pi}{2},\sqrt{1{-}\smash{k^2}})\ .
\end{equation}
In the following we sometimes drop the parameter $k$, i.e.~write
$\sn[u;k]=\sn(u)$ etc.

The Jacobi elliptic functions generalize the trigomonetric functions
and satisfy analogous identities, including
\begin{align}
\sn^{2}u+\cn^{2}u&\=1\ ,\\
k^{2}\sn^{2}u+\dn^{2}u&\=1\ ,\\
\cn^{2}u+\sqrt{1{-}\smash{k^2}}\,\sn^{2}u&\=1
\end{align}
as well as
\begin{align}
\sn[u;0]&\=\sin u\ ,\\
\cn[u;0]&\=\cos u\ ,\\
\dn[u;0]&\=1\ .
\end{align}

One may also define $\cn$, $\dn$ and $\sn$ as solutions $y(x)$ to the
respective differential equations
\begin{align}
y''&\=(2{-}k)^{2}y+y^{3}\ , \\
y''&\=-(1{-}2k^{2})y+2k^{2}y^{3}\ , \\
y''&\=-(1{+}k^{2})y+2k^{2}y^{3}\ .
\end{align}


\begin{thebibliography}{99}

\bibitem{GSW}
  M.B.~Green, J.H.~Schwarz and E.~Witten,
  {\it Superstring theory},\\
  Cambridge University Press, Cambridge, 1987.

\bibitem{Corrigan:1982th}
  E.~Corrigan, C.~Devchand, D.B.~Fairlie and J.~Nuyts,
  ``First order equations for gauge fields in spaces of dimension
    greater than four,''
  Nucl.\ Phys.\ B {\bf 214} (1983) 452.

\bibitem{Ward84}
  R.S.~Ward,
  ``Completely solvable gauge field equations in dimension
    greater than four,''\\
  Nucl.\ Phys.\ B {\bf 236} (1984) 381.

\bibitem{DUY}
  S.K.~Donaldson,
  ``Anti-self-dual Yang-Mills connections on a complex algebraic surface
    and stable vector bundles,''
  Proc.\ Lond.\ Math.\ Soc.\ {\bf 50} (1985) 1;\\
  ``Infinite determinants, stable bundles and curvature,''
  Duke Math.\ J.\ {\bf 54} (1987) 231;

 K.K.~Uhlenbeck and S.-T.~Yau,
  ``On the existence of Hermitian-Yang-Mills connections on stable bundles
    over compact K\"ahler manifolds,''
  Commun.\ Pure Appl.\ Math.\ {\bf 39} (1986) 257;\\
  ``A note on our previous paper,''
  {\sl ibid.} {\bf 42} (1989) 703.

\bibitem{CSC}
  M.~Mamone~Capria and S.M.~Salamon,
  ``Yang-Mills fields on quaternionic spaces,''\\
  Nonlinearity {\bf 1} (1988) 517;

  R.~Reyes~Carri\'on,
  ``A generalization of the notion of instanton,''
  Diff.\ Geom.\ Appl.\ {\bf 8} (1998) 1.

\bibitem{Bau}
  L.~Baulieu, H.~Kanno and I.M.~Singer,
  ``Special quantum field theories in eight and other dimensions,''
  Commun.\ Math.\ Phys.\ {\bf 194} (1998) 149
  [arXiv:hep-th/9704167].

\bibitem{Tian}
  G.~Tian,
  ``Gauge theory and calibrated geometry,''\\
  Ann.\ Math.\ {\bf 151} (2000) 193
  [arXiv:math/0010015 [math.DG]];

  T.~Tao and G.~Tian,
  ``A singularity removal theorem for Yang-Mills fields in higher dimensions,''
  J.\ Amer.\ Math.\ Soc.\ {\bf 17} (2004) 557.

\bibitem{DT}
  S.K.~Donaldson and R.P.~Thomas,
  ``Gauge theory in higher dimensions,''\\
  in: {\it The Geometric Universe},
  Oxford University Press, Oxford, 1998;

  S.K.~Donaldson and E.~Segal,
  ``Gauge theory in higher dimensions II'',\\
  arXiv:0902.3239 [math.DG].

\bibitem{Popov}
  A.D.~Popov,
  ``Non-Abelian vortices, super-Yang-Mills theory and Spin(7)-instantons,''\\
  Lett.\ Math.\ Phys.\ {\bf 92} (2010) 253
  [arXiv:0908.3055 [hep-th]];

  D.~Harland and A.D.~Popov,
  ``Yang-Mills fields in flux compactifications on homogeneous manifolds
  with SU(4)-structure,''
  arXiv:1005.2837 [hep-th]. 

\bibitem{group1}
  D.B.~Fairlie and J.~Nuyts,
  ``Spherically symmetric solutions of gauge theories in eight dimensions,''
  J.\ Phys.\ A {\bf 17} (1984) 2867;

  S.~Fubini and H.~Nicolai,
  ``The octonionic instanton,''
  Phys.\ Lett.\ B {\bf 155} (1985) 369;

  T.A.~Ivanova and A.D.~Popov,
  ``Self-dual Yang-Mills fields in $d{=}7, 8$, octonions and Ward equations,''
  Lett.\ Math.\ Phys.\  {\bf 24} (1992) 85;\\
  ``(Anti)self-dual gauge fields in dimension $d{\ge}4$,''
  Theor.\ Math.\ Phys.\ {\bf 94} (1993) 225.

\bibitem{IL}
  T.A.~Ivanova and O.~Lechtenfeld,
  ``Yang-Mills instantons and dyons on group manifolds,''\\
  Phys.\ Lett.\ B {\bf 670} (2008) 91
  [arXiv:0806.0394 [hep-th]].

\bibitem{ILPR}
  T.A.~Ivanova, O.~Lechtenfeld, A.D.~Popov and T.~Rahn,
  ``Instantons and Yang-Mills flows on coset spaces,''
  Lett.\ Math.\ Phys.\  {\bf 89} (2009) 231
  [arXiv:0904.0654 [hep-th]];

 T.~Rahn,
  ``Yang-Mills equations of motion for the Higgs sector of SU(3)-equivariant
    quiver gauge theories,''
  arXiv:0908.4275 [hep-th].

\bibitem{HILP}
D.~Harland, T.A.~Ivanova, O.~Lechtenfeld and A.D.~Popov,\\
  ``Yang-Mills flows on nearly K\"ahler manifolds and $G_2$-instantons,''
  arXiv:0909.2730 [hep-th].

\bibitem{group3}
  M.~Grana,
  ``Flux compactifications in string theory: A comprehensive review,''\\
  Phys.\ Rept.\ {\bf 423} (2006) 91
  [arXiv:hep-th/0509003];

  M.R.~Douglas and S.~Kachru,
  ``Flux compactification,''\\
  Rev.\ Mod.\ Phys.\ {\bf 79} (2007) 733
  [arXiv:hep-th/0610102];

  R.~Blumenhagen, B.~Kors, D.~L\"ust and S.~Stieberger,
  ``Four-dimensional string compactifications with D-branes, orientifolds
    and fluxes,''
  Phys.\ Rept.\ {\bf 445} (2007) 1
  [arXiv:hep-th/0610327].

\bibitem{hetold}
  A.~Strominger,
  ``Superstrings with torsion,''
  Nucl.\ Phys.\ B {\bf 274} (1986) 253;

  C.M.~Hull,
  ``Anomalies, ambiguities and superstrings,''
  Phys.\ Lett.\ B {\bf 167} (1986) 51 (1986);\\
  ``Compactifications of the heterotic superstring,''
  Phys.\ Lett.\ B {\bf 178} (1986) 357 (1986);

  D.~L\"ust,
  ``Compactification of ten-dimensional superstring theories over Ricci flat
    coset spaces,''\\
  Nucl.\ Phys.\ B {\bf 276} (1986) 220;

  B.~de Wit, D.J.~Smit and N.D.~Hari Dass,
  ``Residual supersymmetry of compactified D=10 supergravity,''
  Nucl.\ Phys.\ B {\bf 283} (1987) 165.

\bibitem{But}
  J.-B.~Butruille,
  ``Homogeneous nearly K\"ahler manifolds'',
  arXiv:math/0612655 [math.DG];

  F.~Xu,
  ``SU(3)-structures and special lagrangian geometries,''
  arXiv:math/0610532 [math.DG].

\bibitem{group4}
  A.~Tomasiello,
  ``New string vacua from twistor spaces,''\\
  Phys.\ Rev.\ D {\bf 78} (2008) 046007 [arXiv:0712.1396 [hep-th]];

  C.~Caviezel, P.~Koerber, S.~Kors, D.~L\"ust, D.~Tsimpis and M.~Zagermann,\\
 ``The effective theory of type IIA AdS4 compactifications on nilmanifolds 
  and cosets'',\\
 Class.\ Quant.\ Grav.\ {\bf 26} (2009) 025014
 [arXiv:0806.3458 [hep-th]];

A.D.~Popov,
  ``Hermitian-Yang-Mills equations and pseudo-holomorphic bundles
    on nearly K\"ahler and nearly Calabi-Yau twistor 6-manifolds,''\\
  Nucl.\ Phys.\  B {\bf 828} (2010) 594
  [arXiv:0907.0106 [hep-th]].

\bibitem{BPST}
A.A.~Belavin, A.M.~Polyakov, A.S.~Schwarz and Y.S.~Tyupkin,\\
  ``Pseudoparticle solutions of the Yang-Mills equations,''
  Phys.\ Lett.\ B {\bf 59} (1975) 85.

\bibitem{Raj}
R.~Rajaraman, {\it Solitons and instantons}, North-Holland, Amsterdam, 1984.

\bibitem{MS}
N.~Manton and P.~Sutcliffe, {\it Topological solitons},\\
   Cambridge University Press,  Cambridge, 2004.

 \bibitem{group5}
J.-X.~Fu, L.-S.~Tseng and S.-T.~Yau,
``Local heterotic torsional models,''\\
 Commun.\ Math.\ Phys.\  {\bf 289} (2009) 1151
  [arXiv:0806.2392 [hep-th]];

 M.~Becker, L.-S.~Tseng and S.-T.~Yau,
``New heterotic non-K\"ahler geometries,''\\
  arXiv:0807.0827 [hep-th];

K.~Becker and S.~Sethi,
  ``Torsional heterotic geometries,''\\
   Nucl.\ Phys.\  B {\bf 820} (2009) 1
[arXiv:0903.3769 [hep-th]].

  \bibitem{group6}
  I.~Benmachiche, J.~Louis and D.~Martinez-Pedrera,\\
  ``The effective action of the heterotic string compactified on manifolds with
  SU(3) structure,''\\
  Class.\ Quant.\ Grav.\  {\bf 25} (2008) 135006
  [arXiv:0802.0410 [hep-th]];

  M.~Fernandez, S.~Ivanov, L.~Ugarte and R.~Villacampa,\\
``Non-K\"ahler heterotic string compactifications with non-zero fluxes and
constant dilaton,''\\
Commun.\ Math.\ Phys.\  {\bf 288} (2009) 677
[arXiv:0804.1648 [math.DG]];

G.~Papadopoulos,
``New half supersymmetric solutions of the heterotic string,''\\
Class.\ Quant.\ Grav.\  {\bf 26} (2009) 135001
  [arXiv:0809.1156 [hep-th]];

H.~Kunitomo and M.~Ohta,
  ``Supersymmetric AdS$_3$ solutions in heterotic supergravity,''\\
Prog.\ Theor.\ Phys.\  {\bf 122} (2009) 631
  [arXiv:0902.0655 [hep-th]].

\bibitem{group7}
G.~Douzas, T.~Grammatikopoulos and G.~Zoupanos,
  ``Coset space dimensional reduction and Wilson flux breaking of
  ten-dimensional N=1, E(8) gauge theory,''\\
  Eur.\ Phys.\ J.\  C {\bf 59} (2009) 917
  [arXiv:0808.3236 [hep-th]];

A.~Chatzistavrakidis and G.~Zoupanos,
  ``Dimensional reduction of the heterotic string over nearly-K\"ahler
  manifolds,''
  JHEP {\bf 09} (2009) 077
  [arXiv:0905.2398 [hep-th]];

A.~Chatzistavrakidis, P.~Manousselis and G.~Zoupanos,
  ``Reducing the heterotic supergravity on nearly-K\"ahler coset spaces,''
  Fortsch.\ Phys.\  {\bf 57} (2009) 527
  [arXiv:0811.2182 [hep-th]].

\bibitem{KN}
  S.~Kobayashi and K.~Nomizu,
  {\it Foundations of differential geometry}, vol.1,\\
  Interscience Publishers, 1963.

\bibitem{Kub}
  Yu.A.~Kubyshin, I.P.~Volobuev, J.M.~Mourao and G.~Rudolph,\\
 ``Dimensional reduction of gauge theories, spontaneous compactification
 and model building,''\\
    Lect.\ Notes Phys.\  {\bf 349} (1990) 1.

\bibitem{KZ}
  D.~Kapetanakis and G.~Zoupanos,
  ``Coset space dimensional reduction of gauge theories,''\\
  Phys.\ Rept.\  {\bf 219} (1992) 1.

\bibitem{LPS}
O.~Lechtenfeld, A.D.~Popov and R.J.~Szabo,
``Quiver gauge theory and noncommutative vortices,''
  Prog.\ Theor.\ Phys.\ Suppl.\  {\bf 171} (2007) 258
  [arXiv:0706.0979 [hep-th]];\\
``SU(3)-equivariant quiver gauge theories and nonabelian vortices,''\\
  JHEP {\bf 08} (2008) 093
  [arXiv:0806.2791 [hep-th]].

\bibitem{CS}
  S.~Chiossi and S.~Salamon,
  ``The intrinsic torsion of SU(3) and $G_2$ structures,''\\
  arXiv:math/0202282 [math.DG].

\bibitem{group8}
S.J.~Avis and C.J.~Isham,
  ``Vacuum solutions for a twisted scalar field,''\\
  Proc.\ Roy.\ Soc.\ Lond.\  A {\bf 363} (1978) 581;

N.S.~Manton and T.M.~Samols,
  ``Sphalerons on a circle,''
  Phys.\ Lett.\  B {\bf 207} (1988) 179;

J.Q.~Liang, H.J.W.~M\"uller-Kirsten and D.H.~Tchrakian,\\
  ``Solitons, bounces and sphalerons on a circle,''
  Phys.\ Lett.\  B {\bf 282} (1992) 105.


\end{thebibliography}
\end{document}